\documentclass
{aa}
\newcommand{\sty}{\scriptstyle}
\newcommand{\ssty}{\scriptscriptstyle}
\newcommand{\tsty}{\textstyle}
\newcommand{\be}{\begin{equation}}
\newcommand{\ee}{\end{equation}}
\newcommand{\lb}[1]{\label{#1}}

\newcommand{\obs}[1]{[#1]_{{\tsty {\ssty obs}}}}
\newcommand{\bigobs}[1]{\left[#1\right]_{{\tsty {\ssty obs}}}}
\newcommand{\gen}[1]{#1_{\ssty i}}

\newcommand{\dl}{d_{\ssty L}}
\newcommand{\da}{d_{\ssty A}}
\newcommand{\dg}{d_{\ssty G}}
\newcommand{\gaml}{\gamma_{\ssty L}}

\newcommand{\gamg}{\gamma_{\ssty G}}

\newcommand{\dpr}{d_{\ssty Pr}}

\newcommand{\nc}{n_{\ssty C}}
\newcommand{\dz}{d_z}

\newcommand{\Mbar}{\bar{M}}

\newcommand{\Mv}{{\cal M}_{\ssty v}}
\newcommand{\etal}{et al.\ }
\usepackage{txfonts}
\usepackage{graphicx}

\begin{document}
\title{Relativistic cosmology number densities \\ and the luminosity function}
\author{Alvaro S.\ Iribarrem \inst{1}\thanks{iribarrem@astro.ufrj.br}
\and Amanda R.\ Lopes \inst{1}\thanks{amanda05@astro.ufrj.br}
\and Marcelo B.\ Ribeiro \inst{2}\thanks{mbr@if.ufrj.br}
\and William R.\ Stoeger \inst{3}\thanks{wstoeger@email.arizona.edu}}
\institute{Valongo Observatory, Federal University of Rio de Janeiro,
Ladeira Pedro Ant\^onio, 43, Rio de Janeiro, CEP 20080-090, Brazil
\and
Physics Institute, Federal University of Rio de Janeiro,
CxP 68532, Rio de Janeiro, CEP 21941-972, Brazil
\and
Vatican Observatory Research Group, Steward Observatory,
University of Arizona, Tucson, AZ 85721, USA}
\date{}
\abstract{}
{This paper studies the connection between the relativistic number
density of galaxies down the past light cone in a
Friedmann-Lema\^{\i}tre-Robertson-Walker spacetime with non-vanishing
cosmological constant and the galaxy luminosity function (LF) data. It
extends the redshift range of previous results presented in Albani et al.,
where the galaxy distribution was studied out to $z=1$. Observational
inhomogeneities were detected at this range. This research also searches
for LF evolution in the context of the framework advanced by Ribeiro
and Stoeger, further developing the theory linking relativistic cosmology
theory and LF data.}
{Selection functions are obtained using the Schechter parameters and
redshift parametrization of the galaxy luminosity functions obtained
from an I-band selected dataset of the FORS Deep Field galaxy survey
in the redshift range $0.5\le z \le5.0$ for its blue bands and $0.75
\le z \le 3.0$ for its red ones. Differential number counts, densities
and other related observables are obtained, and then used with the
calculated selection functions to study the empirical radial distribution
of the galaxies in a fully relativistic framework.}
{The redshift range of the dataset used in this work, which is up
to five times larger than the one used in previous studies, shows an
increased relevance of the relativistic effects of expansion when
compared to the evolution of the LF at the higher redshifts. The results
also agree with the preliminary ones presented in Albani et al.,
suggesting a power-law behavior of relativistic densities at high
redshifts when they are defined in terms of the luminosity distance.}
{}
\titlerunning{Relativistic cosmology number densities and the LF}
\authorrunning{A.S.\ Iribarrem et al.\ }
\maketitle

\section{Introduction}

The galaxy volume number density, that is, the number
of galaxies enclosed in a given volume, is a very important quantity
in cosmology. It gives information about the density of the
mass-energy in the universe and its evolution, allowing us to
test various observational features of cosmological models like,
for instance, the observational inhomogeneities of the number
counts data down our past light cone (Albani et al.\ 2007). This
important quantity can be observationally determined by a careful
analysis of data from galaxy redshift surveys, as has
been done in a systematic way by several authors who derived
the galaxy luminosity function (LF) $\phi$ from such
surveys. The LF of galaxies is a number density per unit of
luminosity (Peacock 1999). {The LF stemming
from the galaxy distribution in a given dataset\footnote{In
this paper the term LF refers to the observationally determined
quantity. It must be noted, however, that the LF does not depend
only on observations, but also on an assumed cosmology. Thus,
although it is a way of presenting observations, it also contains
theoretical information. It could be considered to be fully
observational \textit{if} the assumed cosmology is observationally
well-substantiated.}} is commonly fitted using the profile due
to Schechter (1976), 
\begin{equation}
\lb{schechter}
\phi (l) = \phi^\ast \, l^{\alpha} \, e^{-l},
\end{equation}
where $l \equiv L/L_{\ssty *}$, $L$ being the observed luminosity,
$L_{\ssty *}$ the luminosity scale parameter, $\phi^\ast$ a
normalization parameter and $\alpha$ the faint-end slope parameter.
{Various papers on the LF have used a variety of datasets in
different wavelength regions:}
Lin \etal (1999), Fried \etal (2001), Blanton \etal (2003),
Pozzetti \etal (2003), Bell \etal (2003), Norman \etal (2004),
Wilmer \etal (2006), Ly \etal (2007), and Tzanavaris \&
Georgantopoulos (2008) obtained Schechter-type LF parameters for
galaxies with redshift values out to $z \sim 1.5$. Poli \etal (2003),
and Rudnick \etal (2003) did the same for galaxies with redshift
values out to $z \sim 3$. More recently Bouwens \etal (2007), and
Gabasch \etal (2008) obtained those parameters for galaxies with
redshift values out to $z \sim 6$. It was found that the {LF
evolves with redshift} in all galaxy redshift surveys. 

Using the LF data to study observational features of cosmological
models requires, however, some model linking relativistic cosmology
number count theory with LF astronomical data and practice. One
approach to this theoretical link was developed by Ribeiro and Stoeger
(2003; hereafter RS03), who started from very general relativistic
considerations and then specialized their theoretical results, applying
them to the LF data provided by the CNOC2 redshift survey (Lin \etal
1999). This provides an example of how such a link is established, as
well as facilitating various consistency tests between the data and the
expected number count predictions from the assumed cosmological model.
In a sequel Albani \etal (2007; hereafter A07) used the same CNOC2
dataset to actually obtain and analyze various types of observational
relativistic densities, that is, the ones defined on the observer's
past light cone.\footnote{Usually a density is just a density,
and the adjective ``relativistic'' does not apply. However, in
this paper this expression has to do with the different ways the density can be
defined, depending on the different measures of distance we use,
since in relativistic cosmology a distance can be defined by
different measures. In this context the adjective ``cosmological''
could also be used. See \S \ref{dis} and \ref{densi} below.} 

The aim of this work is to further study the observed number densities
of galaxies down the observer's past light cone. We want to focus on
observational inhomogeneities of the number count data, analyzing
these with differential density and integral density measures in order
to separate out the effects of cosmic expansion from the effect of LF
evolution. As has been discussed in detail in A07 and in Rangel Lemos \&
Ribeiro (2008), these measures are capable of revealing observational
inhomogeneities even in spatially homogeneous cosmological models. This
is because spatial homogeneity is defined on constant time
hypersurfaces, and is therefore a built-in feature of Friedmann
cosmologies, whereas observational homogeneity is defined by
measures of mass-energy densities remaining unchanged along the past
light cone. We also seek to investigate if there is significant
evidence of power-law behavior in the observed galaxy distribution at
high values of redshift, since A07 detected such a behavior out to
$z \approx 1$. {The importance of power-laws comes from the fact
that they are indicative of self-organized criticality in dynamical
systems. In other words, scale invariant phenomena such as power laws
are emergent features of a distribution, this being a mechanism by
which complexity arises in nature through simple local interactions.}

The study presented here expands the results presented in RS03 and
{A07}, since the model specific equations obtained in RS03 were
applicable only to LF data whose parameters assumed the Einstein-de
Sitter (EdS) cosmology. And although the study presented in A07 was
not limited to the EdS model, it was limited to the CNOC2 survey,
which only probed out to $z=0.75$ and extrapolated to
$z=1$. In addition, although both papers made some important
theoretical connections between the relativistic cosmology equations and
the LF data for a specific cosmology, they omitted important steps.
{These can be summarized as follows. The equations
and methodology used in A07 for dealing with non-zero cosmological
constant models are generalized, fully presented and discussed in
this paper. An appendix containing a detailed algorithm on how to
obtain the key results is also included. The role of the completeness
function $J$ connecting theory and observations, as advanced in
RS03, is made explicit here, adding clarity to the results and
presenting them in a more straightforward manner. Finally, a
comparison of the geometrical versus evolutionary effects, the latter
being empirically obtained from the LF, on the relativistic volume
number densities is discussed, considering a broad spectral
classification of late versus early type galaxies}. So, this work
further advances the treatment of both RS03 and A07 and presents a
comprehensive and detailed discussion of how one can proceed in
applying the very general, model independent, relativistic cosmology
equations to specific cosmologies, thus linking \textit{any}
cosmological model to LF data. This is necessary if we want to use
LF data to observationally test more general cosmological models
(any more general models will be inhomogeneous and/or anisotropic),
a task we intend to carry out in forthcoming papers. {Table
\ref{novelties} summarizes the new contributions of this paper as
compared with the results obtained in the previous ones}.

{In this paper we assume a
Friedmann-{Lema\^{\i}tre}-Robertson-Walker (FLRW) cosmology
with non-zero cosmological constant $\Lambda$ and derive model
independent relativistic equations which enable us to solve
numerically expressions which include LF data.} The numerical
scheme is presented in detail and applied to a galaxy redshift survey
dataset that assumes the FLRW standard cosmological model in its volume
and distance definitions: the I-band selected luminosity functions of
Gabasch \etal (2004 and 2006; hereafter G04 and G06 respectively) for
the FORS Deep Field (FDF). From the Schechter parameters and their
redshift parametrization appearing in G04 and G06, we calculated
the selection functions $\psi$ in eight bandwidths and in equally spaced
redshift intervals of 0.25 spanning the range $0.5\le z \le 5.0$, for the
blue bands (G04), and $0.75 \le z \le 3.0$ for the red bands (G06). Once
the selection functions were calculated, we were able to obtain
the observational number counts, differential number counts and
differential and integral densities, as defined in RS03 and Ribeiro
(2005) and further discussed in A07 and Rangel Lemos \& Ribeiro (2008).
These quantities were then used to study observational inhomogeneities in
the relativistic radial distribution of galaxies belonging to that
dataset. Therefore, the calculations of this paper go to {much higher
redshift} than the ones presented in A07. The results reached
here generally agree with theirs, also indicating that observational
inhomogeneities in the relativistic galaxy distributions can 
arise due to geometrical, null-geodesic effects, even in the context
of spatially homogeneous universes (Ribeiro 1992, 1995, 2001, 2005;
Rangel Lemos \& Ribeiro 2008).

The plan of the paper is as follows. In \S \ref{theory} we briefly
present the FLRW model, basically to set up notation and write essential
results. Then we apply the general, model independent, equations of
relativistic cosmology number-count theory to this universe model
{in order to obtain the differential equations governing
the evolution of the two basic quantities of our analysis, namely, the
scale factor and the relativistic number counts. \S \ref{nprob}
describes in detail our numerical approach for solving these two
differential equations, as well as our procedures for
deriving all other quantities relevant in our analysis. In doing that
we also explain how they connect to the the scale factor and the
number counts.} In \S \ref{obs} we further develop general equations
connecting LF data with relativistic cosmology number-count theory
and obtain the selection functions for the FDF dataset. In \S
\ref{inom} we study the behavior of radial statistics of the
relativistic densities obtained in the context of the FLRW model
with $\Lambda\not=0$ and out to $z=5.0$. We present our conclusions
in \S \ref{conclusions}. {The appendix presents an algorithm
detailing how the methodology discussed in \S \ref{theory}, \S
\ref{nprob} and \S \ref {obs} can be applied to a given
parametrized LF in order to obtain the observational relativistic
densities discussed in \S \ref{inom}.}

\section{Standard cosmology with non-zero cosmological constant}
\label{theory}

\subsection{The scale factor}\label{scale}
We begin by writing the FLRW line element as follows,
\begin{equation}
\label{metric}
ds^2 = - \; c^2dt^2 + S^2 \left[ \frac{dr^2}{1-kr^2} + r^2 \left(
       d\theta^2 + \sin^2 \theta \; d\phi^2 \right) \right],
\end{equation}
where the time-dependent function $S=S(t)$ is the cosmic scale
factor, $k$ is the curvature parameter $(k=+1, 0, -1)$ and
$c$ is the light speed. As is well known, the Einstein's field
equations with the metric corresponding to this line element yields
the Friedmann equation,
which, if the cosmological constant $\Lambda$ is included,
may be written ({\it e.g.\ }Roos 1994),
\begin{equation}
\label{friedini}
H^2 = \frac{8 \pi G \rho_m}{3} + \frac{\Lambda}{3} - \frac{k c^2}{S^2},
\end{equation}
where $\rho_m$ is the  matter density and we have assumed the
usual definition for the Hubble parameter, 

\begin{equation}
\label{Hdef}
H(t) \equiv \frac{1}{S(t)} \frac{d S(t)}{dt}.
\end{equation}
Let us now define the vacuum energy density in terms of the
cosmological constant, 
\begin{equation}
\label{rhoLdef}
\rho_{\Lambda} \equiv \frac{\Lambda}{8\pi G}.
\end{equation}
Since the critical density at the present time is given by,
\be
{\rho_{0,c} \equiv \frac{3 {H_0}^2}{8 \pi G}},
\lb{rhoc}
\ee
where $H_0$ is the  Hubble constant, the following
relative-to-critical density parameter relations hold,
\begin{equation}
\label{omega0}
\Omega_0 \equiv \Omega_{m_0} + \Omega_{\Lambda_0} = {\frac{\rho_0}{\rho_{0,c}} =
\frac{\rho_{m_0}}{\rho_{0,c}} + \frac{\rho_{\Lambda_0}}{\rho_{0,c}}}. 
\end{equation}
We have used the zero index to indicate observable quantities at the
present time. Notice that since $\Lambda$ is a constant, then
$\rho_{\Lambda} = \rho_{\Lambda_0}$. Considering the definitions
(\ref{omega0}), we can rewrite the Friedmann equation (\ref{friedini}) at
the present time as follows,
\begin{equation}
\label{k}
kc^2 = {H_0}^2 {S_0}^2 (\Omega_0 - 1).
\end{equation}
In addition, from the law of conservation of energy applied to the zero
pressure, matter dominated era, we know that,
\be
\rho_m \propto S^{-3} \; \; \Rightarrow \; \; \rho_{m_0} \propto {S_0}^{-3},
\ee
{which leads to,
\begin{equation}
\label{vscale}
\frac{\rho_m}{\rho_{m_0}} = \frac{{S_0}^3}{S^3} \; \; \Rightarrow \; \;
\Omega_m = \Omega_{m_0} \frac{{S_0}^3{H_0}^2}{S^3 H^2}, 
\end{equation}
since the matter-density parameter can also be written in terms of
the critical density as,}
\be
\lb{omegam}
\Omega_m = \frac{\rho_m}{\rho_c} = \frac{8 \pi G}{3 {H}^2} \, {\rho_m}.
\ee
{We can rewrite equation (\ref{friedini}) as a first order ordinary
differential equation in terms of the scale factor $S(t)$, by using the
results in equations (\ref{Hdef})-(\ref{omegam}), yielding,}
\begin{equation}
\label{SvsT}
\frac{dS}{dt} = H_0 { \left[ \frac{\Omega_{m_0} {S_0}^3}{S} +
                \Omega_{\Lambda_0} {S}^2
		- (\Omega_0-1) {S_0}^2 \right]}^{1/2}.
\end{equation}

{The problem we deal with in this paper requires the solution of the
equation above as well as of another differential equation for the
cumulative number count $N$ (see eq.\ \ref{NvsR} below). The
simplest approach is to simultaneously solve these two differential
equations by numerical means.} 

\subsection{Relativistic number counts} \lb{numcount}

Let us write the completely general, cosmological, model-independent
expression derived by Ellis (1971) for the number count
of cosmological sources, which takes fully into account relativistic
effects, as follows,
\begin{equation}
\label{eleq}
dN = (\da)^2 d\Omega[n(-k^au_a)]_P \; dy.
\end{equation}
Here $dN$ is the number of cosmological sources in a volume
section {at a point P} down the null cone, $n$ is the number
density of radiating sources per unit of proper volume in a section
of a bundle of light rays converging towards the observer and subtending
a solid angle $d\Omega$ at the observer's position, $\da$ is the
area distance of this section from the observer's viewpoint (also
known as angular diameter distance, observer area distance
and corrected luminosity distance), $u^a$ is the observer's 
4-velocity, $k^a$ is the tangent vector along the light rays and
$y$ is the affine parameter distance down the light cone
constituting the bundle (see RS03, \S 2.1, figure 1). The number
density, that is, the number of cosmological sources per proper
volume unit, can be related to the matter density $\rho_m$ by means of, 
\begin{equation}
n = \frac{\rho_m}{{\cal M}_g}.
\lb{n1}
\end{equation}
where, ${\cal M}_g$ is simply the  average galaxy rest mass,
dark matter included.

One should point out that the details of the galaxy mass function
and how it evolves with the redshift are imprinted in the {LF itself.
As a consequence,} it will be included in any selection functions
stemming from the LF. For the empirical
purposes of this paper this equation is correct to an order of
magnitude and should be regarded only as an estimate. This means that
equation (\ref{n1}) enables us to connect the theoretical relativistic
quantities to the LF data and thus include the redshift evolution of
the mass function empirically {(more on this point at the
end of \S \ref{ge} below)}. However, to actually extract from
the LF its implicit galaxy mass function requires some assumption
about the function ${\cal M}_g(z)$. We shall deal with this problem
in a forthcoming paper.

So, for the empirical approach of this paper it is enough to assume
a constant average galaxy rest mass with the working value of
${\cal M}_g \approx 10^{11} {M}_{\sun}$ based on the
estimate by Sparke \& Gallagher (2000). Hence, if we use equations
{(\ref{vscale}) and (\ref{omegam})} we can rewrite equation
(\ref{n1}) as,
\begin{equation}
\label{n}
n = \left( \frac{3 \Omega_{m_0} {H_0}^2 {S_0}^3}{8 \pi G {\cal M}_g}
    \right) \frac{1}{S^3}.
\end{equation}

The past radial null geodesic in the geometry given by metric
(\ref{metric}) may be written 
\be
   \frac{dt}{dr}= - \left( \frac{S}{c \sqrt{1-kr^2}} \right).
   \lb{null}
\ee
Since both coordinates $t$ and $r$ are functions of the affine
parameter $y$ along the past null cone, we have that
\be \frac{dt}{dy}=\frac{dt}{dr} \frac{dr}{dy} = - \left( \frac{S}{c
    \sqrt{1-kr^2}} \right) \frac{dr}{dy}.
    \lb{null2}
\ee
Assuming now that both source and observer are comoving, then 
$u^a=c \; {\delta_0}^a$ and the following results hold, 
\begin{equation}
\label{ng}
-k^au_a = -c \;k^ag_{0a} = - c \; k^0 = - c \frac{dt}{dy} =
         \left(\frac{S}{\sqrt{1-kr^2}}\right) \frac{dr}{dy}.
\end{equation}
In addition, remembering that the area distance $\da$ is defined by
means of a relation between the intrinsically measured
cross-sectional area element $d\sigma$ of the source and the observed
solid angle $d\Omega_0$ (Ellis 1971, 2007; Pleba\'{n}ski \&
Krasi\'{n}ski 2006), we have that
\begin{equation}
\label{da}
(\da)^2 = \frac{d\sigma}{d\Omega} = \frac{S^2 r^2 ( d\theta^2 + \sin^2
\theta \; d\phi^2 )}{(d\theta^2 + \sin^2 \theta \; d\phi^2)} = (Sr)^2.
\end{equation}
We note that for this metric the area distance, a strictly
observational distance definition, equals the proper distance
$\dpr$, a relativistic one. Considering equation (\ref{k}) and
substituting equations (\ref{n}), (\ref{ng}) and (\ref{da}) into
equation (\ref{eleq}), and remembering that here $d\Omega_0=4\pi$,
we are then able to write the number of sources along the past light
cone in terms of the radial coordinate $r$ and in the FLRW model
defined in \S \ref{scale}. This quantity is given by the following
expression, 
\begin{equation}
\label{NvsR}
\frac{dN}{dr} = \left( \frac{3 \; c \; \Omega_{m_0} {H_0}^2
                {S_0}^3}{2 G {\cal M}_g} \right)
	\left[ \frac{r^2}{
	\sqrt{c^2 - {H_0}^2 {S_0}^2 (\Omega_0 - 1) r^2}} \right] .
\end{equation}

\section{Numerical problem}\label{nprob}

{The differential equations for the scale factor $S$ and the
cumulative number count $N$, equations (\ref{SvsT})
and (\ref{NvsR}) respectively, provide the basic quantities necessary here, 
since the expressions which actually connect the relativistic theory
to LF data are built upon them. However, before these two differential
equations can be solved numerically, some further algebraic
manipulations are necessary. Next we shall describe these steps and
how other quantities relevant to our analysis are written in terms
of these two functions.}

{The numerical problem can be summarized as follows. We take
the radial coordinate $r$ as the independent variable in order to
numerically obtain the functions $N(r)$ and $S(r)$ along the past light
cone. Then all other quantities of interest are written in terms of
these two functions, meaning that numerical results for $N(r)$ and
$S(r)$ allow us to straightforwardly obtain $z\:(r)$ and $dN/dz\:(r)$,
as well as various cosmological distances, their derivatives,
observational volumes and number densities. Thus, all quantities used
in our analysis end up being written in terms of the radial coordinate
$r$.}

\subsection{Scale factor}

{A simple and straightforward approach to solving the
differential equations for the scale factor $S$ and the cumulative
number count $N$ is to take the radial coordinate $r$ as the
independent variable and simultaneously integrate them by numerical
means. To do so we shall proceed as follows.}

Equation (\ref{SvsT}) has its time coordinate implicitly defined in
the scale factor, therefore we need to rewrite that differential
equation in such a way that the independent variable becomes explicit.
Considering equation (\ref{k}), the null geodesic (\ref{null}) becomes,
\begin{equation}
\label{TvsR}
\frac{dt}{dr} = - \left[ \frac{S^2}{c^2-{H_0}^2 {S_0}^2 (\Omega_0
                - 1)r^2} \right] ^{\frac{1}{2}}.
\end{equation}
Along the past light cone, we can write the scale factor in terms of
the radial coordinate as
\begin{equation}
\frac{dS}{dr} = \frac{dS}{dt} \frac{dt}{dr}.
\end{equation}
Thus, we are able to rewrite equation (\ref{SvsT}) in
terms of the radial coordinate,
\begin{equation}
\label{SvsR}
\frac{dS}{dr} = - H_0 \left[ \frac{ (\Omega_{\Lambda_0})S^4
		- {S_0}^2(\Omega_0-1)S^2
                + (\Omega_{m_0}{S_0}^3)S
		}
		{c^2-{H_0}^2 {S_0}^2 (\Omega_0
		- 1)r^2} \right]^{\frac{1}{2}}.
\end{equation}
To find solutions for $S(r)$ we must assume numerical values for
$\Omega_{m_0}$, $\Omega_{\Lambda_0}$ and $H_0$. In this paper we
adopt the FLRW model with $\Omega_{m_{\ssty 0}} = 0.3$,
$\Omega_{\Lambda_{\ssty 0}} = 0.7$ and $H_0=70 \; \mbox{km}
\; {\mbox{s}}^{-1} \; {\mbox{Mpc}}^{-1}$.

The two differential equations (\ref{NvsR}) and (\ref{SvsR})
comprise our numerical problem. Solving them simultaneously enables us
to generate tables for $r$, $S$ and $N$. A computer code using the
fourth-order Runge-Kutta method is good enough to successfully carry
out the numerical tasks. The initial conditions $r_0$ and $N_0$ are set
to zero, whereas $S_0$ can be derived considering that as
$r \rightarrow 0$ the spacetime is approximately Euclidean, that
is, $k \approx 0$. This leads, from equation (\ref{null}), to
${ct=-r}$ as well as $S_0=1$.

\subsection{Differential number count}

The redshift $z$ can be written as
\begin{equation}
\label{red}
1+z=\frac{S_0}{S},
\end{equation}
where it is clear that a numerical solution of the scale factor
$S(r)$ immediately gives us the numerical solution for $z(r)$. We
can derive the differential number counts $dN/dz$ by means
of the expression, 
\begin{equation}
\label{dNdz}
\frac{dN}{dz} = \frac{dr}{dz}\frac{dN}{dr},
\end{equation}
with the help of the useful relation
\begin{equation}
\label{drdzdef}
\frac{dr}{dz} = \frac{dr}{dS}\frac{dS}{dz}.
\end{equation}
Numerically, since we build all our quantities using $N(r)$ and $S(r)$,
any derivative with respect to the redshift we wish to evaluate
will be similarly written in terms of the derivative of that quantity
in terms of the radial coordinate. The derivatives in equation
(\ref{drdzdef}) can be taken from definition (\ref{red}) and equation
(\ref{SvsR}), enabling us to write,
\begin{equation}
\label{drdz}
\frac{dr}{dz} = \frac{S^2}{S_{\ssty 0} \, H_{\ssty 0}}
\left[ \frac{ (\Omega_{\Lambda_0})S^4
		- {S_0}^2(\Omega_0-1)S^2
                + (\Omega_{m_0}{S_0}^3)S
		}
		{c^2-{H_0}^2 {S_0}^2 (\Omega_0
		- 1)r^2} \right]^{-\frac{1}{2}}.
\end{equation}
This, together with equation (\ref{NvsR}), allows us to write
equation (\ref{dNdz}) as,
\begin{eqnarray}
\label{dNdz2}
\frac{dN}{dz} &=& \left( \frac{3 \; c \; \Omega_{m_0} H_0 {S_0}^2}{2 G
                {\cal M}_g} \right) \times \nonumber \\
& & \times \left[ \frac{r^2 S^2}{ \sqrt{ 
		(\Omega_{\Lambda_0})S^4 - {S_0}^2
		(\Omega_0-1)S^2 + (\Omega_{m_0}{S_0}^3)S}} \right].
\end{eqnarray}

\subsection{Proper and comoving volumes}\lb{vol}

As discussed in RS03, most cosmological densities obtained from
astronomical observations assume comoving volumes, whereas
densities derived from theory often assume the local, or
proper, volumes. The luminosity function, for instance,
is nowadays always obtained from galaxy catalogues by assuming the
comoving volume in its calculation. So, we are only able to compare
those observationally derived parameters with theory if we carry out
a conversion of volume units. From metric (\ref{metric}) it is
obvious that,
\begin{equation}
\label{convert}
dV_{\ssty Pr} = \frac{S^3}{\sqrt{1-kr^2}} r^2
     dr \sin\theta d\theta d\phi= S^3 dV_{\ssty C},
\end{equation}
This equation clearly defines the conversion factor between these two
volume definitions.

\subsection{Distance measures}\lb{dis}

So far we have used only the area distance $\da$ as distance
definition. However, other cosmological distances can, and will,
be used later. They can be easily obtained from the area distance
by invoking Etherington's reciprocity law (Etherington
1933; Ellis 1971, 2007), 
\begin{equation}
\label{rec}
\dl = (1+z)^2 \da = (1+z) \; \dg.
\end{equation}
Here $\dl$ is the luminosity distance and $\dg$ is the
galaxy-area distance. The latter is also known as 
angular-diameter  distance, transverse comoving distance or
proper-motion distance. A fourth distance will also be
useful later, the redshift distance $d_z$, defined by
the following equation,
\begin{equation}
\label{dz}
d_z = \frac{c \: z}{H_0}.
\end{equation}
Although the reciprocity law is independent of any cosmological
model, the detailed calculations presented in the previous
sections are for the FLRW cosmology only. This means that in deriving
$\da$, $\dl$, $\dg$ and $\dz$ we write the FLRW expression for $\da$
and derive the others using the general, model independent, expression
(\ref{rec}), as well as equation (\ref{dz}). Thus, considering
equation (\ref{da}) for $\da$ and the reciprocity theorem (\ref{rec}),
it is straightforward to write the other cosmological distances in
terms of the scale factor, as follows,
\be \dl = {S_0}^2 \left( \frac{r}{S} \right), \lb{dl} \ee
\be \dg= S_0 \: r, \lb{dg} \ee
\be d_z=\frac{c}{H_0} \left( \frac{S_0}{S} -1 \right). \lb{dzS} \ee

The derivatives of each distance with respect to the redshift still
need to be determined since they will be necessary later. Starting
with the area distance $\da$, they can be easily obtained from
equations (\ref{da}) and (\ref{red}). Then it follows that,
\begin{equation}
\label{diffda}
\frac{d(\da)}{dz} = \frac{dS}{dz} \frac{dr}{dS} \frac{d (\da)}{dr}=
  - \frac{S^2}{S_0} \left[r + S { \left( \frac{dS}{dr} \right) }^{-1}
\right].
\end{equation}
Using equation (\ref{SvsR}), this expression may be rewritten
as,
\begin{eqnarray}
\frac{d(\da)}{dz} & = & \frac{S^2}{S_0} \left[ \frac{S}{H_0} \sqrt{
    \frac{ {c^2-{H_0}^2 {S_0}^2 (\Omega_0 - 1)r^2}}{{(\Omega_{\Lambda_0})S^4
    -{S_0}^2(\Omega_0-1)S^2 +(\Omega_{m_0}{S_0}^3)S} }} \right. \; - \nonumber \\
    & & \biggl. - \; r \biggr]. 
\lb{diffda2}
\end{eqnarray}
The other observational distances can be numerically calculated from
equation (\ref{diffda}) if we consider the reciprocity law (\ref{rec}).
Therefore, we have,
\begin{equation} 
\label{diffdl}
\frac{d(\dl)}{dz} = 2(1+z)\da + (1+z)^2\frac{d(\da)}{dz},
\end{equation}
\begin{equation}
\label{diffdg}
\frac{d(\dg)}{dz} = \da + (1+z)\frac{d(\da)}{dz}.
\end{equation}
These two equations can also be rewritten in terms of the scale factor
if we consider equations (\ref{red}) and (\ref{diffda2}), yielding,
\begin{eqnarray}
\frac{d(\dl)}{dz} & = & S_0 \left[ \frac{S}{H_0} \sqrt{ \frac{ {c^2-{H_0}^2
      {S_0}^2 (\Omega_0 - 1)r^2} }{ {(\Omega_{\Lambda_0})S^4-{S_0}^2
      (\Omega_0-1)S^2 +(\Omega_{m_0}{S_0}^3)S} }} \; + \right. \nonumber \\
       & & + \; \biggl. r \biggr], 
\lb{diffdl2}
\end{eqnarray}
\be
\label{diffdg2}
\frac{d(\dg)}{dz} = \frac{S^2}{H_0} 
		    \sqrt{
		    \frac{
		{c^2-{H_0}^2 {S_0}^2 (\Omega_0 - 1)r^2} 
		         }{
{(\Omega_{\Lambda_0})S^4-{S_0}^2(\Omega_0-1)S^2 +(\Omega_{m_0}{S_0}^3)S}
                         }}\; .
\ee

From now on we shall generically use $\gen{d}$ to indicate any
observational distance, which can be any one of the four cosmological
distances defined above ($i = {\sty A}$, ${\sty G}$, ${\sty L}$,
${\sty Z}$).

\subsection{Differential and integral densities}\lb{densi}

The differential density $\gen{\gamma}$ {gives the rate
of growth in number counts, or more exactly in their density, as one
moves along the observational distance $\gen{d}$. It is defined by}
(Ribeiro 2005; Albani \etal 2007; Rangel Lemos \& Ribeiro 2008),
\begin{equation}
\label{gammadef}
\gen{\gamma}=\frac{1}{4 \pi (\gen{d})^2}\frac{dN}{d(\gen{d})},
\end{equation}
whereas the integrated differential density, or simply
integral density, gives the number of sources
per unit of observational volume located inside the observer's
past light cone out to a distance $\gen{d}$. Is is written as,
\be
\gen{\gamma}^\ast=\frac{1}{\gen{V}} \int\limits_{\gen{V}} \gen{\gamma}
\; d\gen{V},
\lb{gest}
\ee
where $\gen{V}$ is the observational volume, 
\begin{equation}
\lb{volume}
\gen{V}= \frac{4}{3} \pi (\gen{d})^3.
\end{equation}
These quantities are useful in determining whether or not, and within 
what ranges, a spatially homogeneous cosmological model
can or cannot be observationally homogeneous as well (Ribeiro 1995,
2005; Rangel Lemos \& Ribeiro 2008). This is because these densities
behave very differently {depending on the distance measure used
in their definitions --} that is, they show a strong dependence on the
cosmological distance adopted. Therefore, as discussed in Rangel
Lemos \& Ribeiro (2008), {these measures are the ones we
employ in this paper, because they are capable of} probing the possible
observational inhomogeneity of the number counts.

{From} a numerical viewpoint it is preferable to write {the
densities given by equations (\ref{gammadef}) and (\ref{gest}}) in
terms of the redshift. Thus, the differential density (\ref{gammadef})
may be written as,  
\begin{equation}
\label{gamma}
\gen{\gamma}=\frac{dN}{dz}\left\{ 4 \pi (\gen{d})^2 \; \frac{d(\gen{d})}{dz}
         \right\} ^{-1}.
\end{equation}

A final point still needs to be discussed. The integral
density (\ref{gest}) is a result of integrating
$\gen{\gamma}$ over an observational volume. The simplest way of
numerically deriving it is shown in what follows. Let us
differentiate $\gen{\gamma}^\ast$ in terms of the volume,
so that, 
\be
\frac{d \left( \gen{\gamma}^\ast \gen{V} \right)}{d\gen{V}}=
\gen{\gamma}.
\lb{gest1}
\ee
{Considering equation (\ref{gamma}),} this result leads
to the following expression, 
\be
    \frac{d \left( \gen{\gamma}^\ast \gen{V} \right)}{dz}=
    \frac{d \left( \gen{\gamma}^\ast \gen{V} \right)}{d V_i}
    \frac{d V_i}{dz}=\gen{\gamma} \frac{d V_i}{dz}
    = \frac{dN}{dz}.
    \lb{gest2}
\ee
Similarly, it is simple to show that,
\be
  \frac{d \left( \gen{\gamma}^\ast \gen{V} \right)}{dr}=\frac{dN}{dr}.
  \lb{gest2a}
\ee
Finally, from these two equations above, as well as from the
definitions of $\gen{\gamma}$ and $\gen{\gamma}^\ast$, it is easy
to conclude that the following expression holds,
\be \gen{\gamma}^\ast = \frac{N}{\gen{V}}. \lb{gest3} \ee
So, the numerical solution of equation (\ref{NvsR}) together with the
numerical determination of all distances, as given by equations
(\ref{da}), (\ref{dl}), (\ref{dg}), (\ref{dzS}), allow us to calculate
the volume (\ref{volume}) and evaluate $\gamma_{\ssty i}^\ast$.
These results fully determine the numerical problem for the
cosmological model under study.

In conclusion, once $N(r)$ and $S(r)$ are calculated along the past
light cone in terms of the radial coordinate $r$, all other quantities
are straightforwardly obtained with the same functional dependence:
$z\:(r)$, $dN/dz\:(r)$, $d_i\,(r)$, $d(d_i)/dz\:(r)$,
$\gamma_i\:(r)$, $V_i\:(r)$ and $\gamma^\ast_i\:(r)$.   

\section{Observational quantities}\lb{obs}

We shall now specialize the equations above to obtain their
observational counterparts based on the LF data from a specific galaxy
catalog. As a direct consequence, we show how these results can be
used to test the consistency of the number count theory detailed above.

\subsection{General equations}\lb{ge}

Generally speaking, we shall assume that an observational quantity
$\obs{T}$ can be related to its theoretical counterpart $T$ by means
of a completeness function $J$, such that, 
\begin{equation}
\lb{Jdef}
\obs{T} = J \, T.
\end{equation}
RS03 showed that such a completeness function can be obtained by
relating the selection function $\psi$, {which gives
the number of galaxies with luminosity above a given threshold in
a given comoving volume}, to the radial number density
$n_{\ssty C}\,(z)$ in comoving volume units -- that is,
the number of sources in a given comoving volume. As already mentioned, it is
common practice to adopt this volume definition to calculate the
LF from galaxy datasets. The volume number density $n$ obtained
in equation (\ref{n}) comes from the right hand side of Einstein's
field equations {and, therefore, it is written in terms of the
proper volume. Thus, equation (\ref{n}) must have its volume units
corrected by means of equation (\ref{convert}) in order for the
former to be correctly related to a selection function $\psi$
stemming from the LF}.

The relationship allowing us to calculate the completeness
function used in this work may be written,
\be \lb{Jdef2} \psi (z) = J(z) \, n_{\ssty C}(z). \ee
Now, if we go back to equation (\ref{n1}) and the discussion
in the paragraph below it, we can see that applying the
completeness function by means of equation (\ref{Jdef})
actually means replacing the theoretical comoving number density
$n_{\ssty C}$, based on a constant average galaxy rest mass, with
its observed counterpart $\psi$ obtained directly from the LF. In
that sense, all the observational quantities obtained by applying
$J(z)$ to their theoretical counterparts will inherit that same
empirical number count redshift evolution encoded in the
parametrization of the LF. As the goal of this paper is to consider
the effect of redshift evolution on the relativistic number
densities, such an approach should suffice.  Also, in equation
(\ref{Jdef2}) the completeness function is independent of volume
units, therefore, if an observational quantity is obtained using
$J(z)$ by means of equation (\ref{Jdef}) its original volume units
dependence is preserved.  

The selection function may be written in terms of the LF,
\be \lb{psi1} \psi (z) = \int ^{\infty}_{l(z)} \phi(l) dl, \ee
where $l(z)$ is the lower luminosity threshold below which the sources
are not observed. The theoretical radial number density in comoving
volume is given by,
\begin{equation}
\label{radn}
n_{\ssty C} = \frac{N}{V_{\ssty C}} = \frac{3\, N(r)}{4 \, \pi \, r^3}.
\end{equation}

Most observational quantities of interest studied in this paper
require previous knowledge of the observed differential number
counts $\obs{dN}$. Therefore, linking this quantity to its 
theoretical counterpart is an essential step in order to apply
Ellis' equation (\ref{eleq}). In view of this, to find $\obs{dN}$
we identify $T$ as $dN$ in equation (\ref{Jdef}), yielding, 
\begin{equation}
\label{Jdiffzaa}
\bigobs{dN} = J(z) {dN}.
\end{equation}
In essence this expression describes the same number counts as in
equation (\ref{Jdef2}) and both are in agreement with Ellis' (1971)
key equation (\ref{eleq}). Considering now both equations (\ref{Jdef2})
and (\ref{Jdiffzaa}), it easily follows that,
\begin{equation}
\label{Jdiffza}
\bigobs{\frac{dN}{dz}} = J(z) \frac{dN}{dz} = \frac{\psi(z)}{n_{\ssty C}(z)}
       \frac{dN}{dz}.
\end{equation}
This is our key equation relating the relativistic theory to the
observations. The number density of a proper volume is given by
\begin{equation}
\label{radnpr}
n = \frac{N}{V_{\ssty Pr}},
\end{equation}
{whereas the number density of a comoving volume is given by
equation (\ref{radn}).} Thus, we may rewrite equation (\ref{Jdiffza})
as follows,
\begin{equation}
\label{Jdiffzb}
\bigobs{\frac{dN}{dz}} = \frac{V_{\ssty C}}{V_{\ssty Pr}}
       \frac{\psi}{n} \frac{dN}{dz}.
\end{equation}
It is important to point out that the volume transformation in
this equation merely reflects the fact that $J(z)$ is 
independent of volume units, proper or comoving, and does
not in itself change the volume units underlying both
$\bigobs{dN/dz}$ and $dN/dz$ when they are derived by means of
Ellis' equation (\ref{eleq}). It is the number density $n$ in
equation (\ref{eleq}) that defines the volume units of both
$\bigobs{dN/dz}$ and $dN/dz$ (see below).

A further step towards writing these results in terms of the
underlying cosmology is taken if we remember that RS03 showed
that Ellis' (1971) differential number counts (\ref{eleq}) can be
rewritten
\begin{equation}
\frac{dN}{dz} = n \, (\da)^2 \, (1+z) \, d\Omega \, \frac{dy}{dz}.
\lb{eleqb}
\end{equation}
Thus, {equation (\ref{Jdiffzb}) becomes}
\begin{equation}
\lb{A07dNdz2}
\bigobs{\frac{dN}{dz}} =
\left[ \frac{V_{\ssty C}}{V_{\ssty Pr}} (\da)^2 \, (1+z) \,
  d\Omega \; \frac{dy}{dz} \right] \psi .
\end{equation}
Note that in this expression $\bigobs{dN/dz}$ is given in terms of
proper volume units because the number density $n$ is written in terms
of proper volume units. Also note that these expressions are
general and independent of any cosmological model. The specific
cosmology will appear once we specialize the terms inside the
brackets on the right hand side.

Galaxy redshift surveys use bandwidth filters, and they sometimes
include some sort of morphological classification. Therefore,
a selection function $\psi$ which considers a set of filters $W$,
and morphological types $v$, may be written as follows (see A07,
eqs.\ 3-10),
\begin{equation}
\lb{psiWv}
\psi(z) = \sum_{\ssty W} a_{\ssty W} \frac{\sum_{\ssty v} P_{\ssty v}
\, \Mv \: \psi^{\ssty W}_{\ssty v}(z)}{\sum_{\ssty v} P_{\ssty v} \, \Mv}.
\end{equation}
Here $P_{\ssty v}$ is the fraction of galaxies in the dataset
that were classified with the morphological type $v$ (see RS03,
eq.\ 13), $\Mv$ is the typical local rest-mass value for a galaxy of
the morphological type $v$, and $a_{\ssty W}$ are constants introduced
to avoid multiple counting of the same objects in the various filters,
defined as follows (see RS03, eq.\ 46),
\be a_{\ssty W}(z) =1, \; \; \; \; \mbox{for} \; \; \; \;  W=1,
  \lb{aw}
  \ee
and
\be a_{\ssty W}(z) < 1, \; \; \; \; \mbox{for}
    \; \; \; \; W>1.
    \lb{bw}
\ee
{Equation (\ref{bw}) simply states that when there exists more
than one observed waveband ($W>1$), $a_{\ssty W}$ gives the fraction of
galaxies in waveband $W>1$ that are not counted in wavebands
$1,2,\ldots,(W-1)$.} Considering these expressions, the completeness
function defined in equation (\ref{Jdef2}) can be rewritten 
\be
\lb{J}
J(z) = \frac{1}{n_{\ssty C}(z)}
\sum_{\ssty W} a_{\ssty W} \frac{\sum_{\ssty v} P_{\ssty v} \, \Mv
\psi^{\ssty W}_{\ssty v}(z)}{\sum_{\ssty v} P_{\ssty v} \, \Mv}.
\ee
Substituting the expression (\ref{psiWv}) in equation (\ref{A07dNdz2})
we obtain the following result,
\be
\bigobs{\frac{dN}{dz}}  =  \left[ \frac{V_{\ssty C}}{V_{\ssty Pr}} 
(\da)^2 (1+z) d\Omega  \frac{dy}{dz} \right] \sum_{\ssty W} a_{\ssty W} 
\frac{\sum_{\ssty v} P_{\ssty v} \Mv \psi^{\ssty W}_{\ssty v}}
{\sum_{\ssty v} P_{\ssty v} \Mv}.
\lb{eleq2}
\ee
This expression already appeared in A07. Here it has been reached
through a more straightforward and simpler derivation and with the
volume dependence explicitly shown. {This means that volume
conversions like the one discussed above in \S \ref{vol} may be
needed if we are to calculate the observational differential number
count $\obs{dN/dz}$ in a consistent way. The above equation is
shown here only to make explicit how the methodology of 
previous work can be understood in relationship to that being developed in
this paper}. In addition, it is important to note that equations
(\ref{Jdiffza}), (\ref{Jdiffzb}) and (\ref{eleq2}) are {closely
related}. The first two are written in a more compact form, whereas
the last one expands the morphological and bandwidth dependencies
of the selection function and the relativistic features of the
underlying four-dimensional spacetime, as indicated by the
expression (\ref{eleqb}).

{Equation (\ref{eleq2}) also allows us to see how a
theoretical mass evolution does not affect the differential
number counts constructed with LF parameters. This is based
on the form of the summation over $\sty W$, inasmuch as if we
allow an evolution of the galactic mass by considering it
dependent on the redshift, such a dependency will appear both
in the numerator and the denominator of the summation and,
therefore, cancels out, at least to first order. However,
although mass evolution does not directly alter equation
(\ref{eleq2}), it will indirectly change $\obs{dN/dz}$ as the
LF parameters used in its construction will include some form
of source evolution. See \S 2.2 of A07 for more details.}

\subsection{Selection functions} \lb{FORSpsi}

To calculate the completeness function and, as a consequence, other
observational quantities, we need to compute the selection functions
{with respect to redshift} in a given galaxy survey. {G04 and
G06 fitted the LF Schechter parameters} over the redshifts of 5558
I-band selected galaxies in the FORS Deep Field dataset, photometrically
measured down to an apparent magnitude limit of $I_{\ssty AB}=26.8$.
{G04 and G06 also} showed that the selection in the I-band is
expected to miss less than 10\% of the objects detected in the K-band,
given that the AB-magnitudes of the I-band are half a magnitude deeper
than those of the K-band, out to redshift 6, beyond which no signal is
detectable in the I-band due to the Lyman break. Heidt \etal (2001)
also argue that the I-band selection minimizes biases like dust
absorption. All galaxies in those studies were therefore selected in the
I-band and then had their magnitudes for each of the five blue bands
(1500 \AA, 2800 \AA, $u'$, $g'$ and $B$) and the three red ones
($r'$, $i'$ and {$z'$}) computed using the best fitting SED given
by their authors' photometric redshift code convolved with the
associated filter function. {Gabash et al.\ (2004, 2006)
determined} the photometric redshifts by fitting template spectra to
the measured fluxes on the optical and near infrared images of the
galaxies. Heidt \etal (2001) reported that approximately 80\% of the
objects in that sample were classified as Im/Irr-type, 15\% as
E/S0-type or Sa/Sb/Sc-type, and finally 5\% as stars.

The redshift ranges of $0.75 \le z \le 3.0$ for the red bands and
$0.5 \le z \le 5.0$ for the blue bands are large enough for checking
possible observational inhomogeneities, since Rangel Lemos \& Ribeiro
(2008) showed that in the Einstein-de Sitter cosmology it is only
theoretically possible, {i.e., not observationally,} to
distinguish these two features when $z \gtrsim 0.1$, whereas for
$z>0.5$ such a distinction becomes significant. Therefore, using
galaxies whose measured redshifts are mostly greater than $z=1$
allows us to obtain data along the {past} light cone far
enough from our present time hypersurface. One must remember that
spatial homogeneity is defined on our constant time hypersurface 
whereas observational homogeneity occurs along our past light 
cone (Rangel Lemos \& Ribeiro 2008; see also Ribeiro 1992, 1995, 2001,
2005). 

To calculate the actual values of these functions, we begin by writing
the result obtained in RS03 for the limited bandwidth version of the
selection function of a given LF, fitted by a Schechter analytical
profile, in terms of absolute magnitudes,
\begin{eqnarray}
\lb{sf}
\psi^{\ssty W}(z) = 0.4 \ln 10 \, \int_{-\infty}^{M^{\ssty W}_{\ssty lim}(z)}
\phi^\ast(z) 10^{0.4[1+\alpha(z)][M^\ast(z) - \Mbar^{\ssty W}]}
\times \nonumber \\ \times
\exp \{-10^{0.4[M^\ast(z) - \Mbar_{\ssty W}]}\} d\Mbar^{\ssty W},
\end{eqnarray}
where, as discussed earlier, the index $W$ indicates the bandwidth
filter. The equations for the redshift evolution of the LF parameters
adopted by G04 and G06 are,
\begin{eqnarray*}
\phi^{\ast}(z) = \phi^{\ast}_{\ssty 0} \, (1+z)^{B^{\ssty W}}, \\
M^{\ast}(z) = M^{\ast}_0 + A^{\ssty W} \ln (1+z), \\
\alpha(z) = \alpha_{\ssty 0},
\end{eqnarray*}
with $A^{\ssty W}$ and $B^{\ssty W}$ being the evolution parameters
fitted for the different $W$ bands and $M^{\ast}_{\ssty 0}$,
$\phi^{\ast}_{\ssty 0}$ and $\alpha_{\ssty 0}$ the local (z $\approx$ 0)
values of the Schechter parameters as defined in G04 and G06. Since all
galaxies were detected and selected in the I-band, we can write,
\begin{equation}
\label{mlim}
M^{\ssty W}_{\ssty lim}(z) = M^{\ssty I}_{\ssty lim}(z) =
I_{\ssty lim} - 5 \log[\dl (z)] - 25 + A^{\ssty I},
\end{equation}
for a luminosity distance $\dl$ given in Mpc. $I_{\ssty lim}$ is
the limiting apparent magnitude of the I-band of the FDF survey 
and equals to 26.8. Its reddening correction is $A^{\ssty I} =
0.035$ (in Heidt \etal 2001). We computed the selection functions
for all eight bands of the dataset by means of simple numerical
integrations at equally spaced values spanning the whole redshift
interval. The errors were propagated quadratically, as mentioned
before. The results are summarized in tables \ref{selectUV} to
\ref{selectred}. Regarding the blue band dataset of G04, two
distinct patterns of the selection functions in the different bands
are noticeable. The UV bands, 1500 \AA, 2800\AA, and $u'$, evolve
tightly with redshift, having values {that are consistent with
each other within the uncertainties}, while at the same time assuming
values outside the uncertainties of those in the blue optical bands
$g'$ and $B$. Therefore we chose to use the selection functions of
the combined UV bands and those of the combined blue optical bands
separately. The selection functions in the red-band dataset of G06,
$r'$, $i'$ and $z'$ are also combined. {Once the combined
selection functions have been obtained}, we can calculate
$\obs{dN/dz}$ by means of equations (\ref{Jdiffza})
{and (\ref{J})}.

\subsection{Observed number counts}

Following equation (\ref{gest3}), the best way to obtain an observed
relativistic number density $\obs{\gen{\gamma}(z)}$ for a given distance
definition $d_i$ is by calculating the observed number counts
$\obs{N(z)}$. These can be written as,
\begin{equation}
\lb{obsN}
\obs{N(z)} = \int_0^z\bigobs{\frac{dN}{dz'}}dz'.
\end{equation}
The uncertainty of the cumulative number counts $\obs{N}$ can be
derived from the already determined uncertainty in the number counts
$\obs{dN/dz}$. In this regard, the discussion about error analysis
in A07 (Appendix) showed that,
\begin{equation}
\label{deltaN}
\delta \obs{N} = 
\bigobs{\frac{dN}{dz}} \bigobs{\frac{d\,^2N}{dz^2}}^{-1} \;
\delta \bigobs{\frac{dN}{dz}},
\end{equation}
where we can write,
\begin{equation}
\label{d2Ndz2obs}
\bigobs{\frac{d\,^2N}{dz^2}} = \frac{d}{dz}\bigobs{\frac{dN}{dz}} =
J(z) \frac{d\,^2N}{dz^2} + \frac{dJ}{dz}\frac{dN}{dz}.
\end{equation}
The derivative of the completeness function $dJ(z)/dz$ can be
obtained from equation (\ref{Jdef2}) and is proportional to the
derivative of the selection function $\psi^{\ssty W}$ itself with
respect to redshift. This can be directly obtained from the
parametrization of the LF.

Next, we shall use the observed differential number counts
$\obs{dN/dz}$ to study the relativistic radial distribution of the
galaxies in the G04 and G06 datasets in the same manner as in A07.

\section{Relativistic radial statistics of the FDF survey} \lb{inom}

The observational values of the differential number counts
$\obs{dN/dz}$ given in table \ref{dNdztable} for all filters in the
dataset permit us to evaluate the observational differential number
densities $\obs{\gamma}$ for the various cosmological distance
definitions in the FLRW model we are assuming. What it takes to do
so is replacing $dN/dz$ with $\obs{dN/dz}$ in equation (\ref{gamma}),
which is exactly the same as applying the completeness function
$J(z)$ to the theoretical values of the relativistic differential
number densities $\gen{\gamma}$. As discussed at the end of \S
\ref{ge}, such application of the completeness function naturally
includes the observed evolution of the number counts in the
selection functions of a given dataset.

A look at equation (\ref{gamma}) allows us to understand the
separate roles of number counts and geometry, encoded in the term
inside the brackets, in the relativistic differential densities
$\gen{\gamma}$. It shows that the underlying cosmological spacetime
affects such quantities and that their theoretical departure from
homogeneity at high redshifts is ultimately a consequence of the
Universe's expansion, through the inclusion of the scale factor $S$
in the distance definitions, equations (\ref{da}), (\ref{dl}),
and (\ref{dzS}). The absence of the scale factor in the galaxy area
distance $\dg$ (eq.\ \ref{dg}) allows us to identify this distance
with the radial coordinate $r$ in this particular spacetime. As a
consequence, the differential number densities $\gamma_{\ssty G}$
obtained using this distance definition are not subject to the
redshift evolution caused by expansion and will remain constant
if a constant average galaxy rest mass is assumed.

The fact that such differential densities depend on the geometry of
the Universe through the term inside the brackets of equation
(\ref{gamma}) causes them to vary with the redshift differently, and
they do not necessarily remain constant. Such inhomogeneities, however,
are to be understood simply as a consequence of using relativistic
distances, defined on the observer's past light cone. They are
therefore easily reconciled with the Cosmological Principle, which
requires large-scale spatial homogeneity. As can be seen in figures
5 and 6 of A07, both differential and integral number densities
can be inhomogeneous, even when a constant average galaxy mass is
assumed in an otherwise spatially homogeneous FLRW metric. Moreover,
both quantities calculated using $\dg$ remain homogeneous, as expected,
since this distance in this particular spacetime is not affected by
expansion, as discussed before.

We can use this important result to further compare the relativistic
effects of considering densities down the past light cone in an
expanding FLRW spacetime and the observational evolution of the LF
in that same hypersurface. The argument is simple: if we compute the
ratio of the theoretical values of a given relativistic density
at a given redshift -- say the differential number density using the
luminosity distance $d_{\ssty L}$ -- to the one using $\dg$, this
indicates {how much observational inhomogeneity
is introduced through the use of $\dl$ in the calculation of this
relativistic density}. Similarly, the ratio of the observational
value of $\obs{\gamma_{\ssty G}}$ to its theoretical one,
$\gamma_{\ssty G}$, indicates how much observational inhomogeneity
the redshift evolution of the LF introduces. {In fact, considering
equations (\ref{gamma}) and (\ref{Jdiffza}) we can see that such a
ratio is simply the completeness function $J(z)$ in a particular
dataset, that is,
\begin{eqnarray}
\lb{Jratio}
\frac{\obs{\gamma_{\ssty G}}}{\gamma_{\ssty G}} & = &
\frac{\obs{dN/dz}\{ 4 \pi (\dg)^2 \; d(\dg)/dz\}^{-1}}
{dN/dz\{ 4 \pi (\dg)^2 \; d(\dg)/dz\}^{-1}} \nonumber \\ & = & 
\frac{\obs{dN/dz}}{dN/dz}=J(z).
\end{eqnarray}
Now, if there were no significant departure from homogeneity at a
given redshift such ratios should be approximately unity.
Therefore, if we subtract the values of those ratios at a given
redshift from one, this number should express how much inhomogeneity
that relativistic density shows at that redshift: the bigger the
number, the bigger the inhomogeneity. Table \ref{ratios} summarizes
these results. At a fixed redshift entry one can compare the
effects on the homogeneity of the relativistic differential
densities caused by the LF redshift evolution of the different
filters in the G04 and G06 datasets.}

{The results in Table \ref{ratios} allow us to see that
the evolution of the LF in the combined UV bands produces a bigger
departure from homogeneity of the differential densities than
the one in the combined optical bands, and this departure is even bigger
than the one in the combined red ones. On the other hand, fixing
a given ratio, say the observational-to-theoretical ratio for the
UV bands, $\obs{\gamma_{\ssty G}}/\gamma_{\ssty G}$, one can observe
that the departure from homogeneity of these relativistic differential
densities is quite a common feature and tends to increase with
redshift. Finally, by comparing the values of the purely theoretical
$\gamma_{\ssty L}/\gamma_{\ssty G}$ ratio to the various observational-to-theoretical
ones, one can investigate which contribution dominates
at each redshift. That is particularly interesting, since the redshift
evolution of any of the other differential densities can be obtained
by a simple combination of those two separate effects, that of
the expansion of the geometry in the
distance definition and that of the redshift evolution of the LF.
For instance, let us consider the observational differential density
in the combined UV filters calculated using the luminosity
distance $\dl$, that is, $[\gamma_{\ssty L}]_{\ssty UV}$. Its
departure from homogeneity can be investigated using the ratio
$[\gamma_{\ssty L}]_{\ssty UV}/\gamma_{\ssty G}$, as discussed above.
Remembering that, by construction $[\gamma_{\ssty L}]_{\ssty UV}=
J_{\ssty UV} \, \gamma_{\ssty L}$, it follows,
\be
\frac{[\gamma_{\ssty L}]_{\ssty UV}}{\gamma_{\ssty G}}=
\frac{J_{\ssty UV} \, \gamma_{\ssty L}}{\gamma_{\ssty G}}=
\frac{[\gamma_{\ssty G}]_{\ssty UV}}{\gamma_{\ssty G}}
\frac{\gamma_{\ssty L}}{\gamma_{\ssty G}},
\ee
which is simply the combination of the observational-to-theoretical
ratio for that dataset, namely the LF of the combined UV bands with
the purely theoretical ratio for that distance definition, in that
case, $\dl$. The values in Table \ref{ratios} indicate that the
geometrical effect is comparable to the evolution of the LF in the
whole redshift range of the G04 and G06 datasets.}

{At this point, one should note that} determining the spacetime
geometry of the Universe or its density parameters is not the goal
of the present paper. We simply assume the
$\Lambda$CDM metric, a necessary step in calculating the above
mentioned geometrical terms, in order to see how the empirical evolution
of the differential number counts $\obs{dN/dz}$ affects the
theoretical results discussed above.

Similarly, we can obtain the observational values of the
integral densities $\obs{\gamma}^{\ast}$ by means of equation
(\ref{gest2}), or (\ref{gest3}), because, by their very definitions,
we have that $\obs{N}=V_{\ssty C} \; \psi(z)$.

\begin{figure}
\includegraphics[width=10cm]{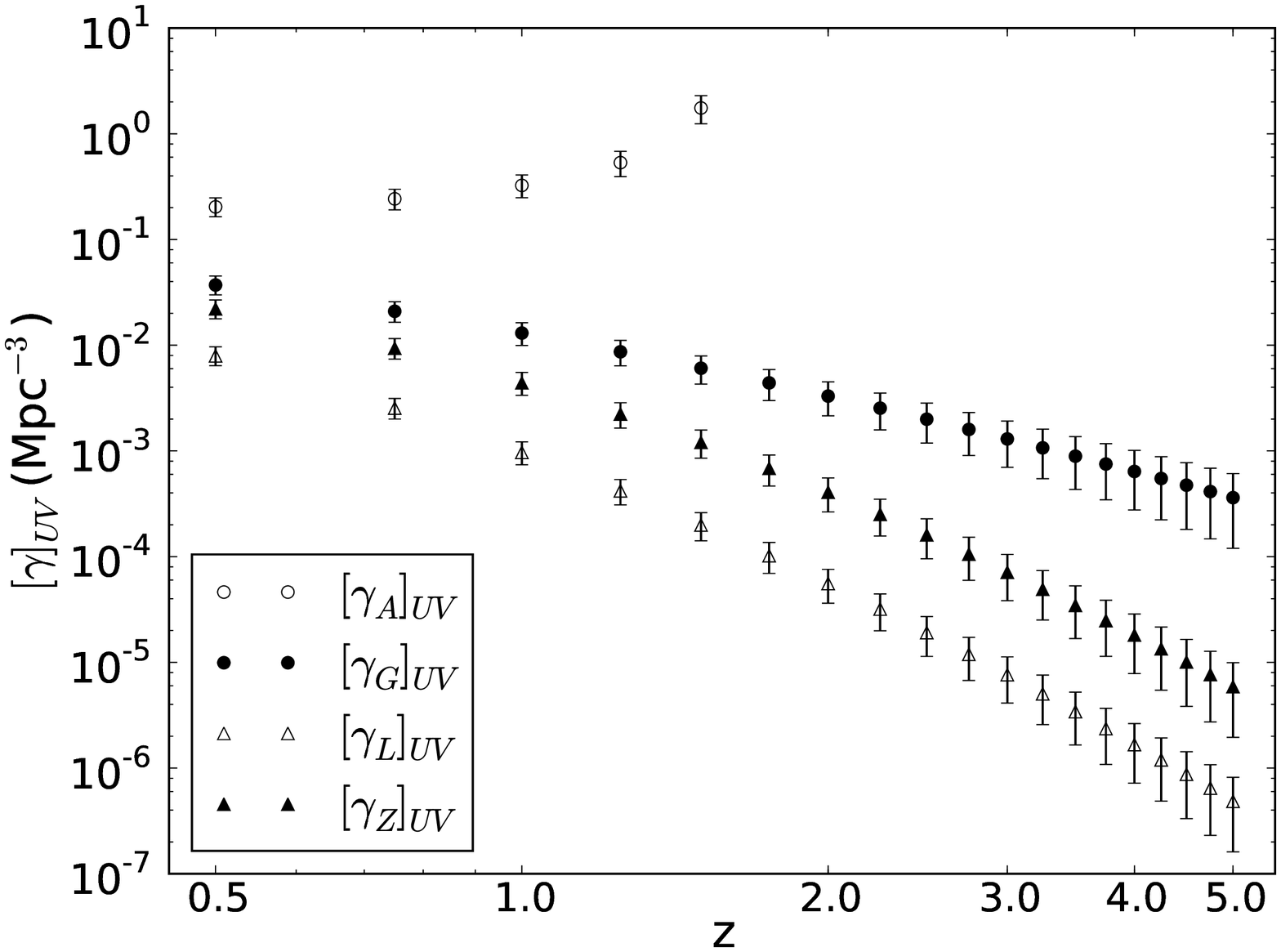}
\caption{{Observational} relativistic differential densities {versus
redshift} in the combined UV bands of the FDF dataset of G04. Symbols
are as in the legend.}
\label{UVgammaplot}
\end{figure}

\begin{figure}
\includegraphics[width=10cm]{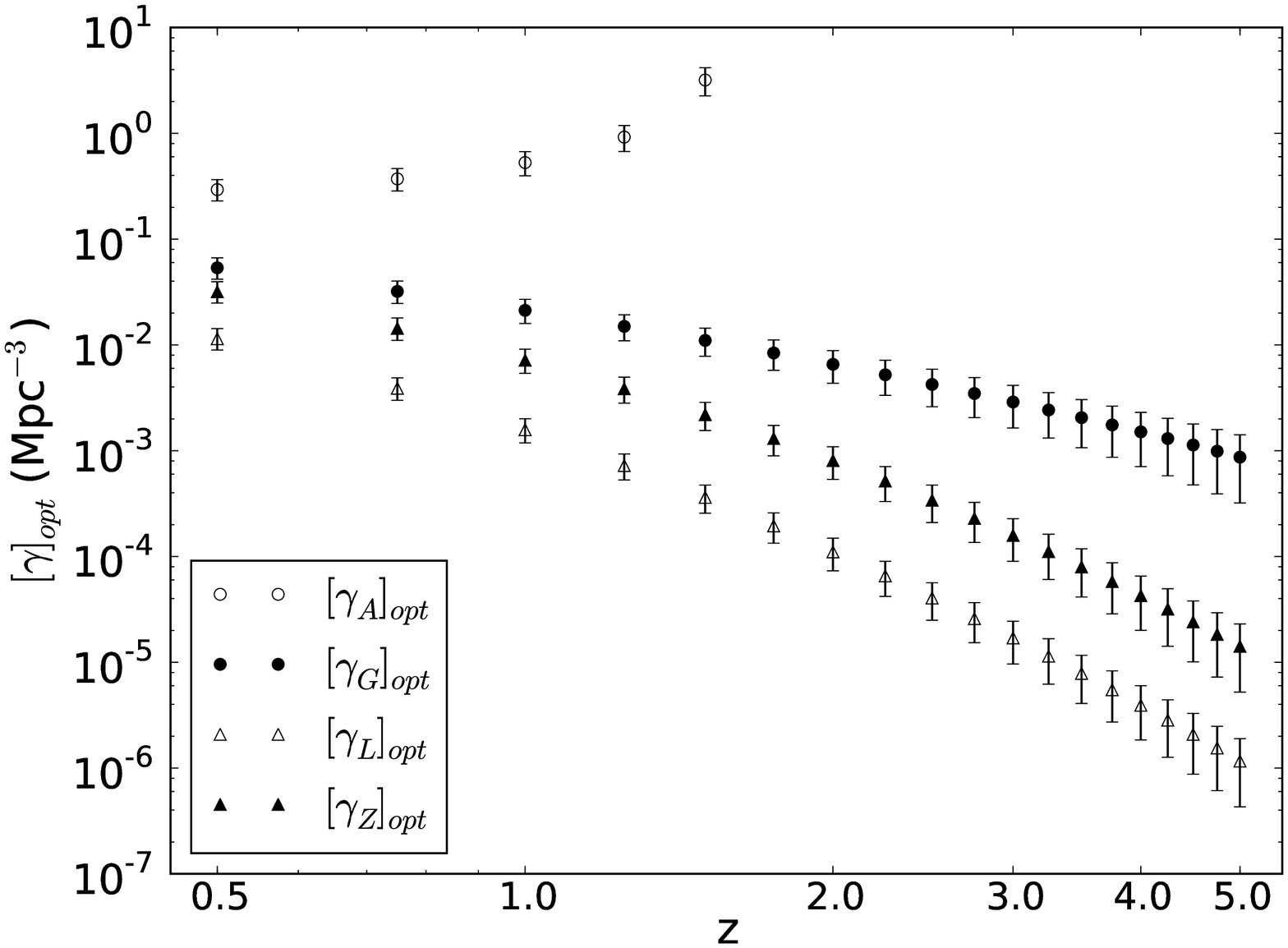}
\caption{{Observational} relativistic differential densities
{versus redshift} in the combined optical bands of the FDF dataset
of G04. {Symbols are as in the legend}.}
\label{Optgammaplot}
\end{figure}

\begin{figure}
\includegraphics[width=10cm]{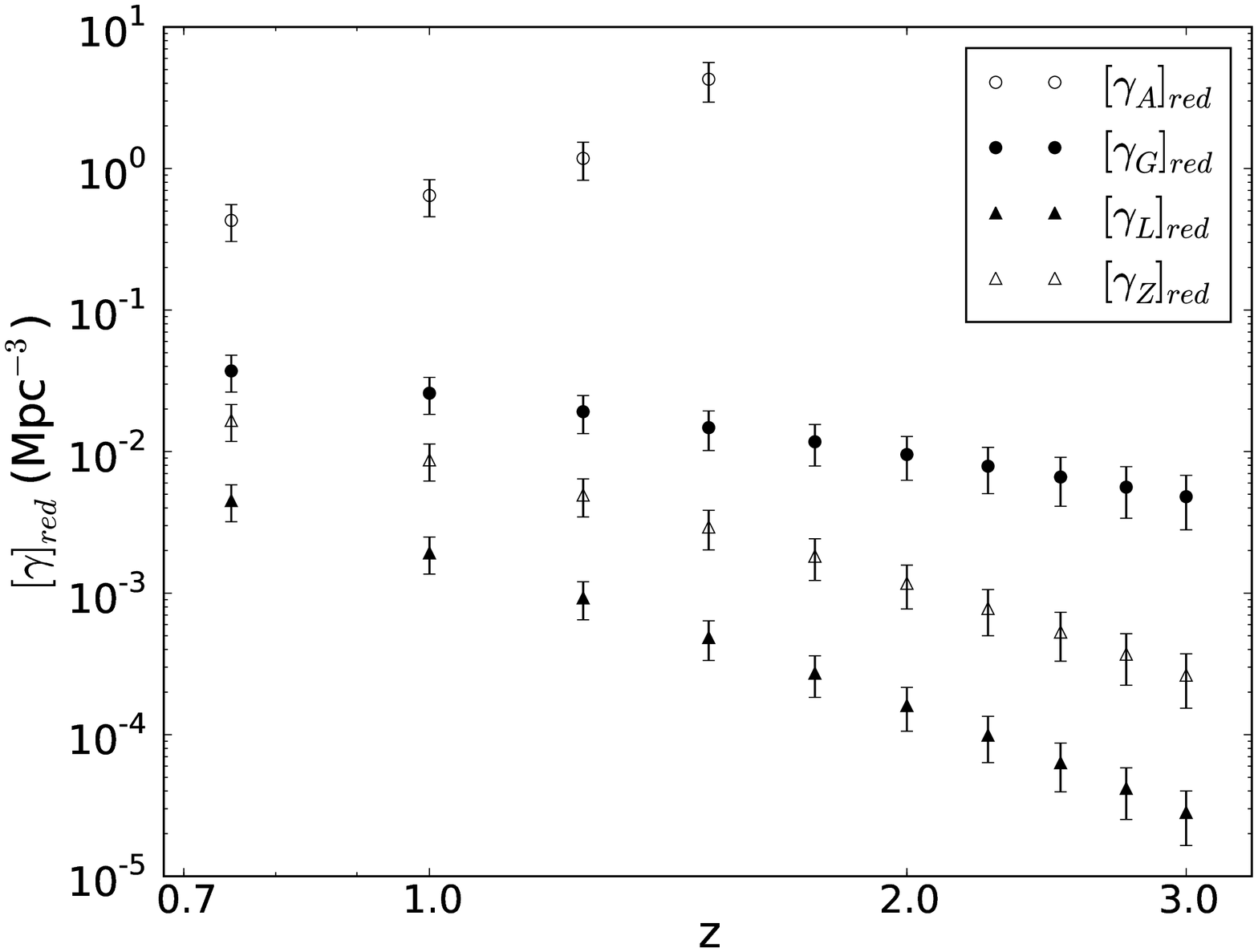}
\caption{{Observational} relativistic differential densities
{versus redshift} in the combined red bands of the FDF dataset
of G06. Symbols are as in the legend.}
\label{redgammaplot}
\end{figure}

Figures \ref{UVgammaplot} and \ref{Optgammaplot} show graphs of the
observational differential densities determined using the combined UV
and optical bands in the G04 dataset plotted against the redshift,
whereas figure \ref{redgammaplot} shows the results for those same
densities versus the redshift in the combined red bands. The
dependence of such relativistic densities on the distance definition,
a known theoretical result (Ribeiro 2005; A07), can be easily observed
in these three figures. In addition, the dependence of
$\obs{\gamma_{\ssty G}}$ with the redshift in these same figures is
solely due to the evolution of the mass function, since its relativistic
volume definition remains constant in the assumed FLRW spacetime. We note
that values of $\obs{\gamma_{\ssty G}}$ obtained from the LF in the
combined red bands (figure \ref{redgammaplot}) seem less dependent on
the redshift than those in the UV and optical bands (figures
\ref{UVgammaplot} and \ref{Optgammaplot}).

\begin{figure}
\includegraphics[width=10cm]{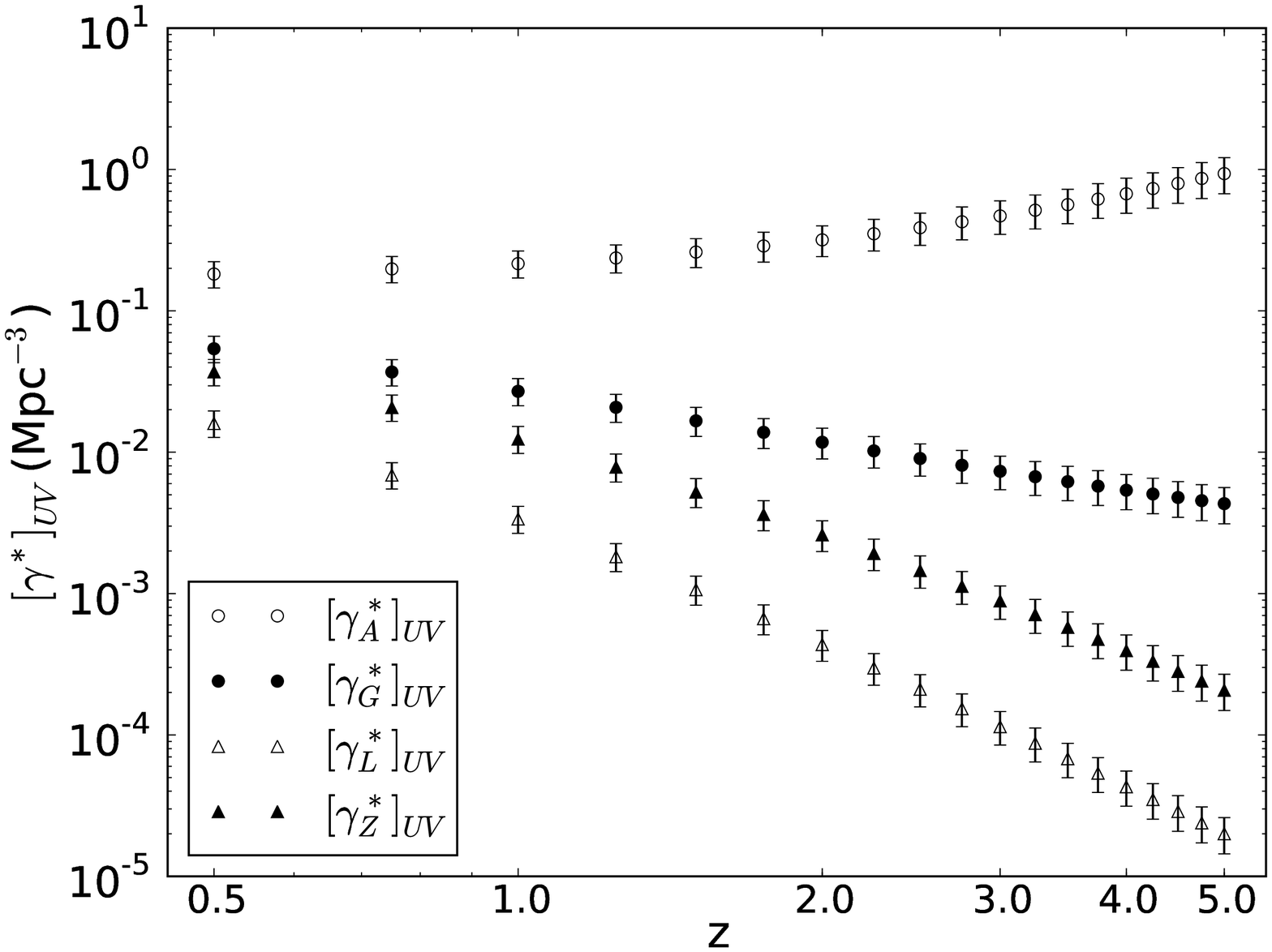}
\caption{{Observational} relativistic integral densities
{versus redshift} in the combined UV bands of the
FDF dataset of G04. Symbols are as in the legend.}
\label{UVgstarplot}
\end{figure}

\begin{figure}
\includegraphics[width=10cm]{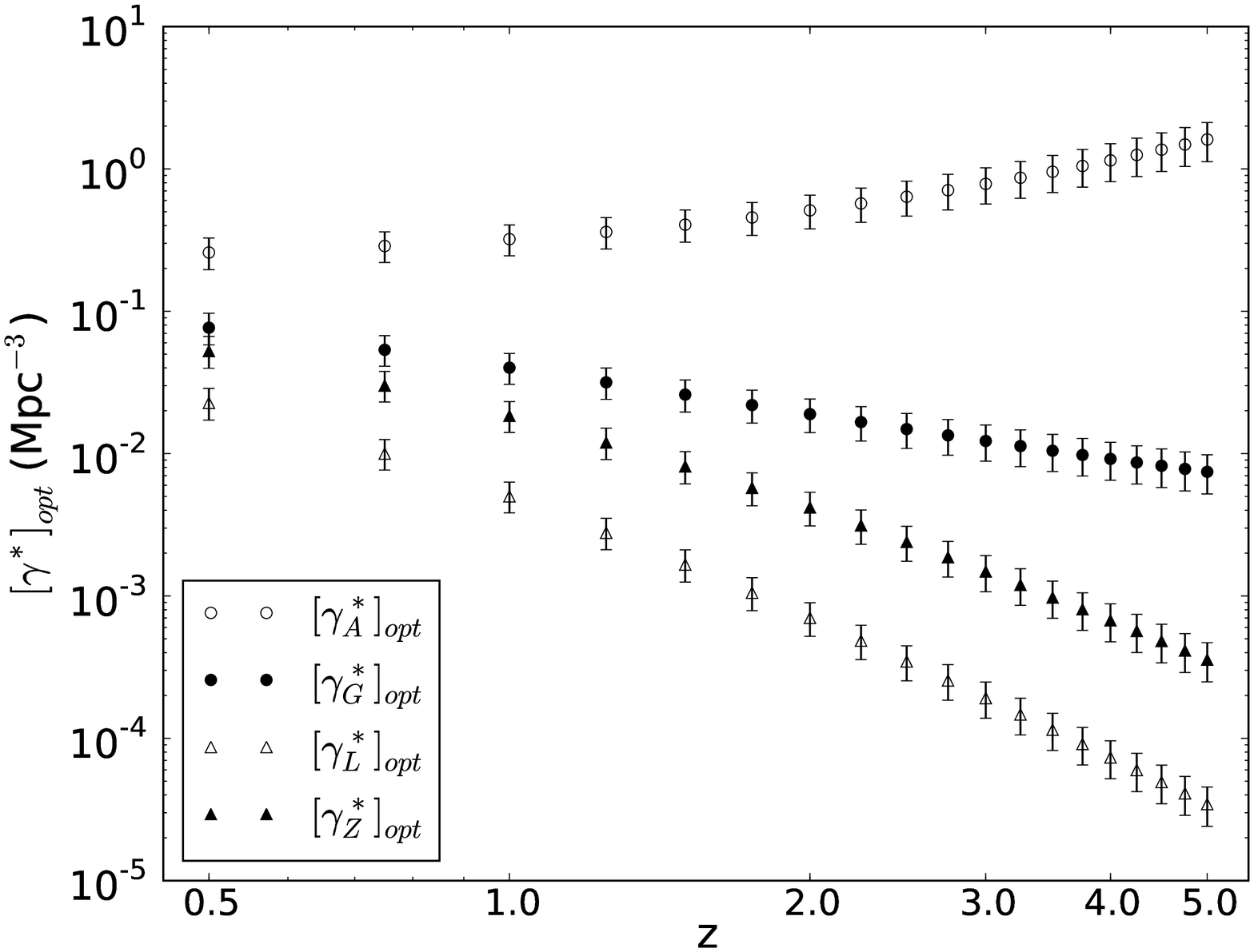}
\caption{{Observational} relativistic integral densities
{versus redshift} in the combined optical bands
of the FDF dataset of G04. {Symbols are as in the legend}.}
\label{Optgstarplot}
\end{figure}

\begin{figure}
\includegraphics[width=10cm]{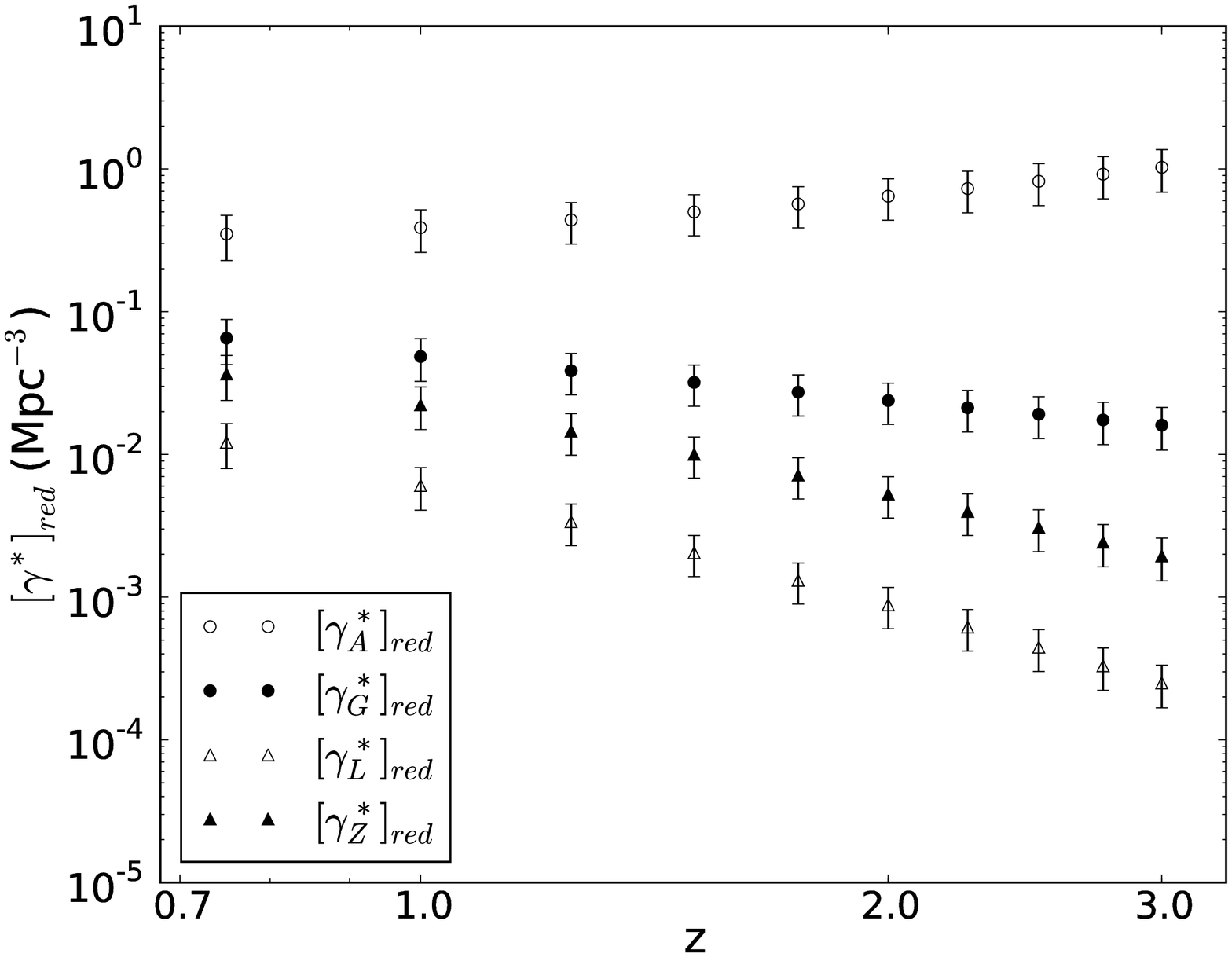}
\caption{{Observational} relativistic integral densities {
versus redshift} in the combined red bands of the FDF dataset of G06.
{Symbols are as in the legend}.}
\label{redgstarplot}
\end{figure}

Similar graphs for the observational integral densities in the UV,
optical and red combined bands versus the redshift are respectively
shown in figures \ref{UVgstarplot}, \ref{Optgstarplot}, and
\ref{redgstarplot}, where their dependency with the distance
definition can also be clearly seen. Although less pronounced in these
smoothed out relativistic densities, the effect of the evolution of
the mass function can still be observed in these three graphs, as
$\obs{\gamma^{\ast}_{\ssty G}}$ is also unaffected by the geometrical
effects of considering the densities on the past null light cone of
an FLRW spacetime. In other words, the LF redshift evolution seems
to further enhance the inhomogeneity of these densities, making
observationally inhomogeneous even the theoretically homogeneous
$\gamma_{\ssty G}$ and $\gamma^\ast_{\ssty G}$. We observe that in
figure \ref{redgstarplot}, $\obs{\gamma^{\ast}_{\ssty G}}$ is also
less dependent on the redshift than in the UV and optical bands 
shown in figures \ref{UVgstarplot} and \ref{Optgstarplot}.

\begin{figure}
\includegraphics[width=10cm]{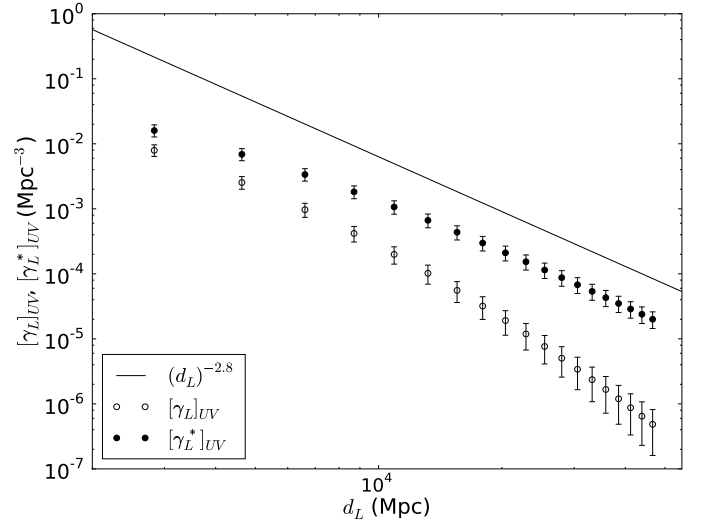}
\caption{{Observational} relativistic differential and integral
densities versus luminosity distance in the combined UV bands of the
FDF dataset of G04. The solid line was drawn just as reference, to
show the apparent power-law behavior of both relativistic densities
in that distance definition at high redshifts. {Symbols are as 
in the legend}.}
\label{UVdlplot}
\end{figure}

\begin{figure}
\includegraphics[width=10cm]{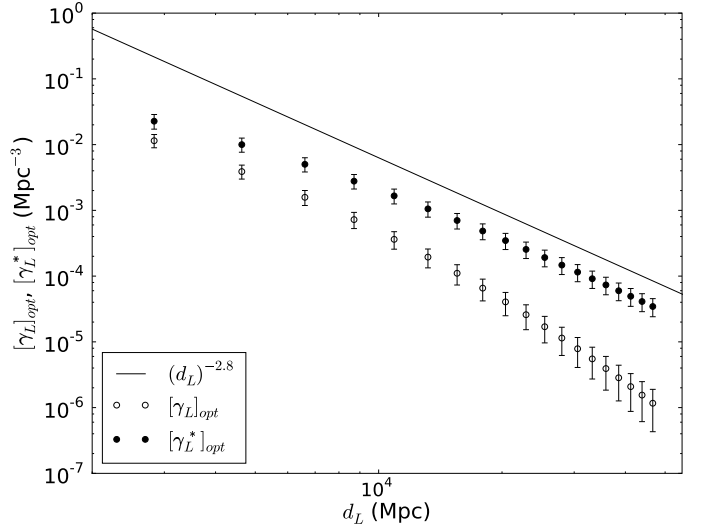}
\caption{Observational relativistic differential and integral
densities versus luminosity distance in the combined optical bands 
of the FDF dataset of G04. The solid line is plotted just as a
reference to the power law behavior stated in the legend. Symbols
are as in the legend.}
\label{Optdlplot}
\end{figure}

\begin{figure}
\includegraphics[width=10cm]{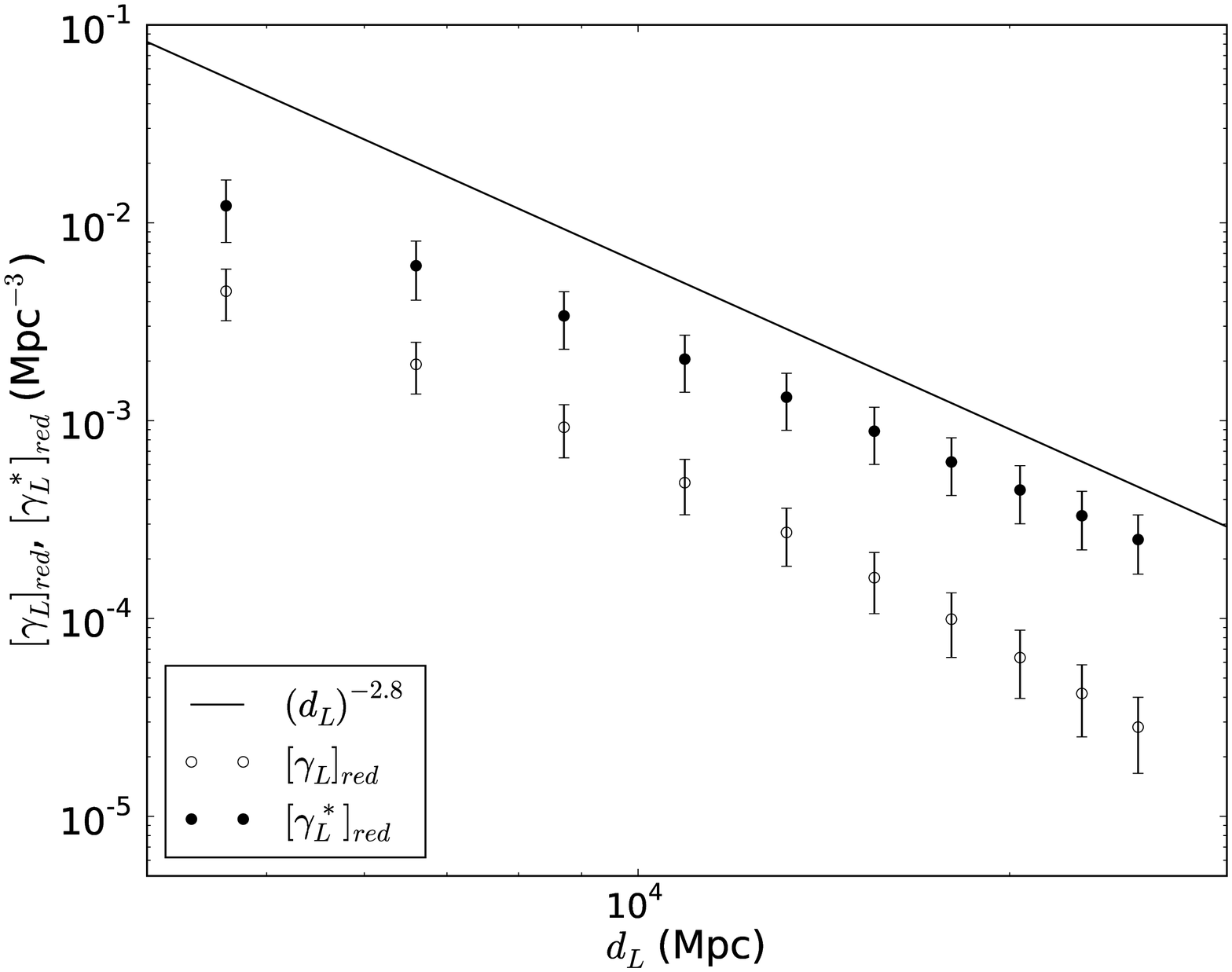}
\caption{Observational relativistic densities versus luminosity 
distance in the combined red bands of the FDF dataset of G06. 
The solid line is shown as a reference to the power law behavior
stated in the legend. Symbols are as in the legend.}
\label{reddlplot}
\end{figure}

Figures \ref{UVdlplot}, \ref{Optdlplot}, and \ref{reddlplot}
plot the observational differential and integral densities 
built with the luminosity distance in the combined UV, optical and
red bands, respectively, versus the luminosity distance. A power-law
pattern for both densities in the three combined bands can be
observed when one compares the data points with the solid
line drawn just as reference in all three plots. The slope of
the reference lines is the same in all bands, which indicates that
there is little change in the slope of the data points.
This possibly indicates that such an effect
is predominantly relativistic. 

\begin{figure}
\includegraphics[width=10cm]{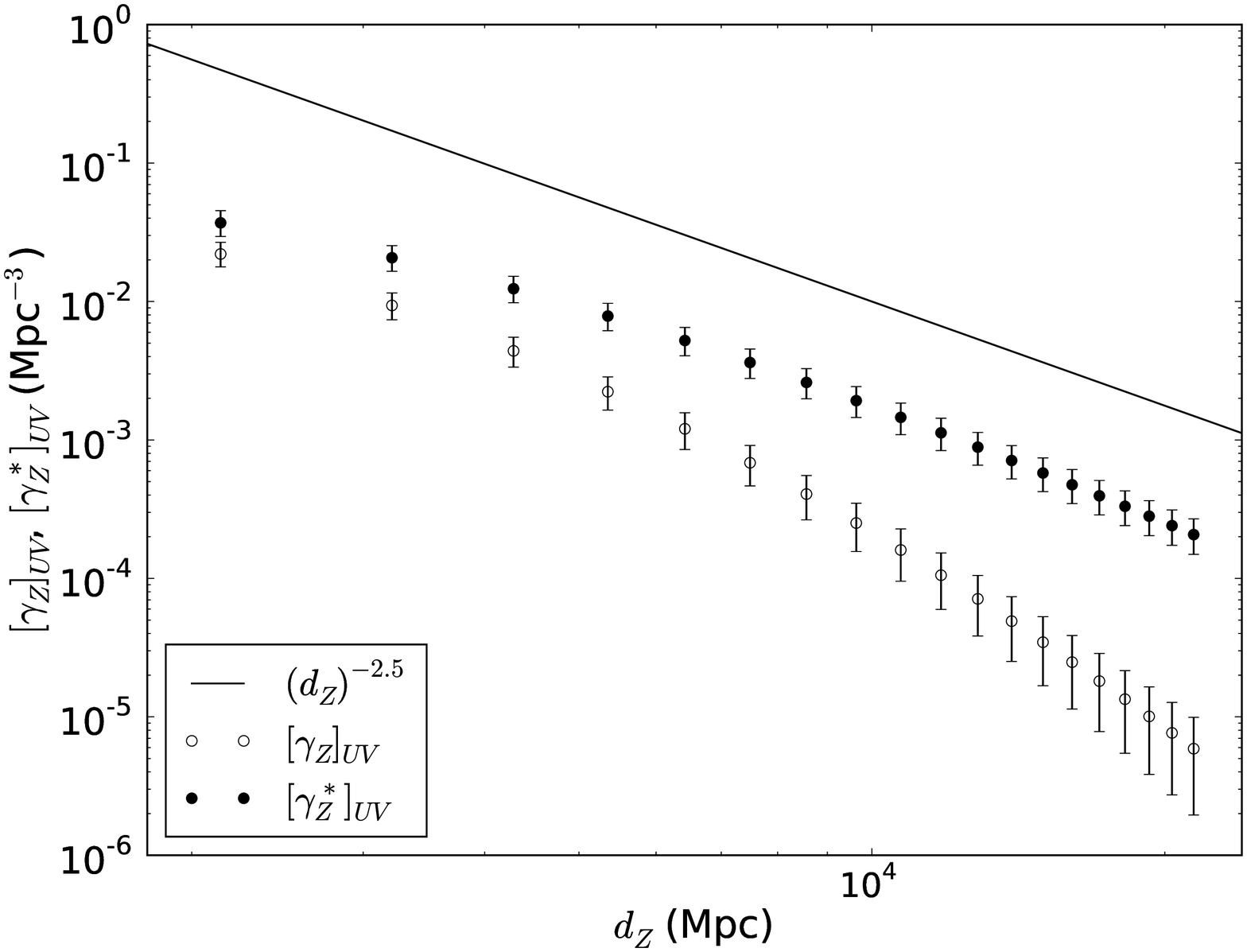}
\caption{{Observational} relativistic differential {and
integral} densities versus redshift distance in the combined UV bands 
of the FDF dataset of G04. The solid line is shown just as a reference
to the power law behavior stated in the legend. Symbols are as in the
legend.}
\label{UVdzplot}
\end{figure}

\begin{figure}
\includegraphics[width=10cm]{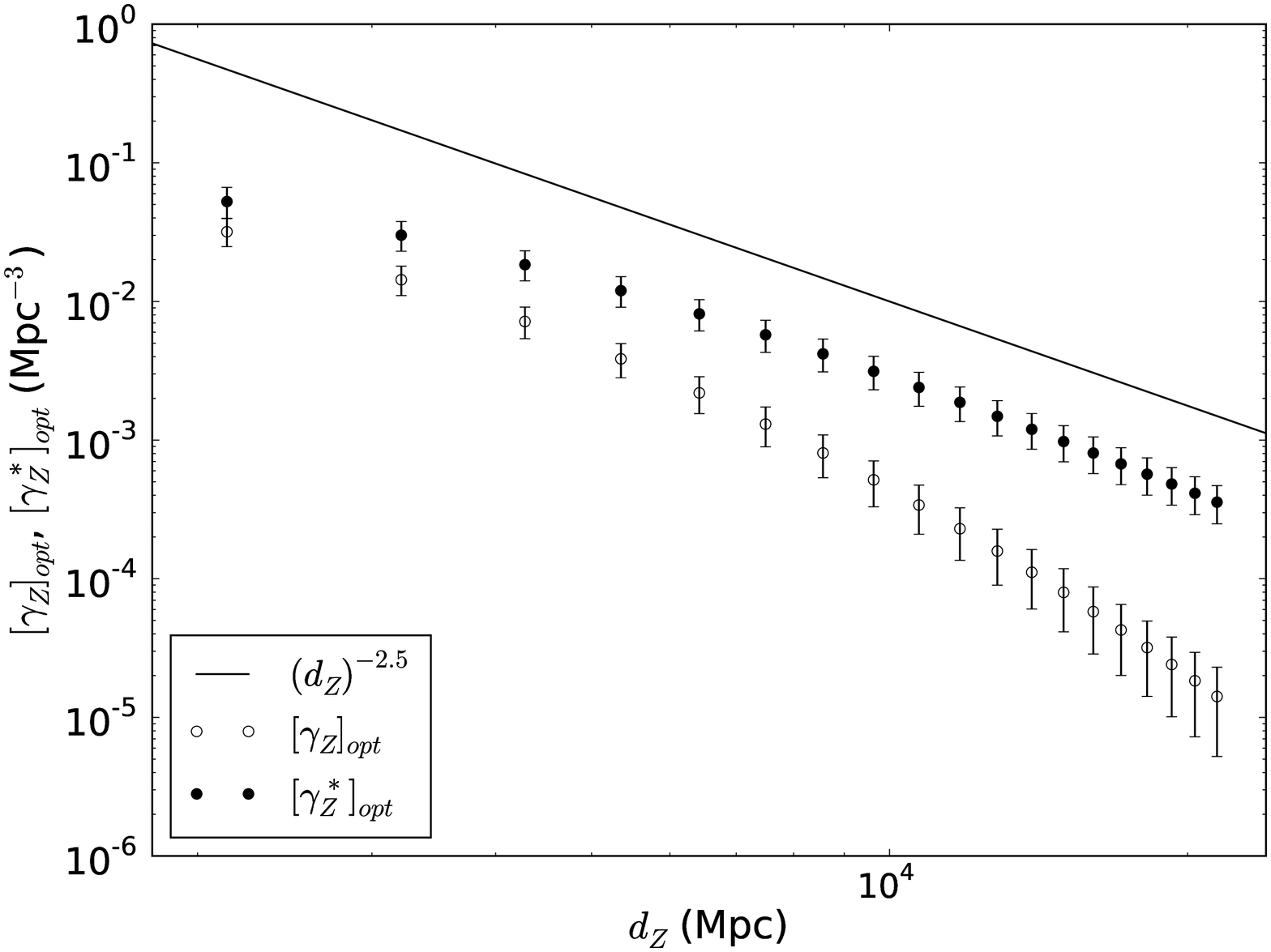}
\caption{{Observational} relativistic differential and integral
densities versus redshift distance in the combined optical bands of 
the FDF dataset of G04. The solid line is shown as a reference to
the power law behavior stated in the legend. Symbols are as in the
legend.} \label{Optdzplot}
\end{figure}

\begin{figure}
\includegraphics[width=10cm]{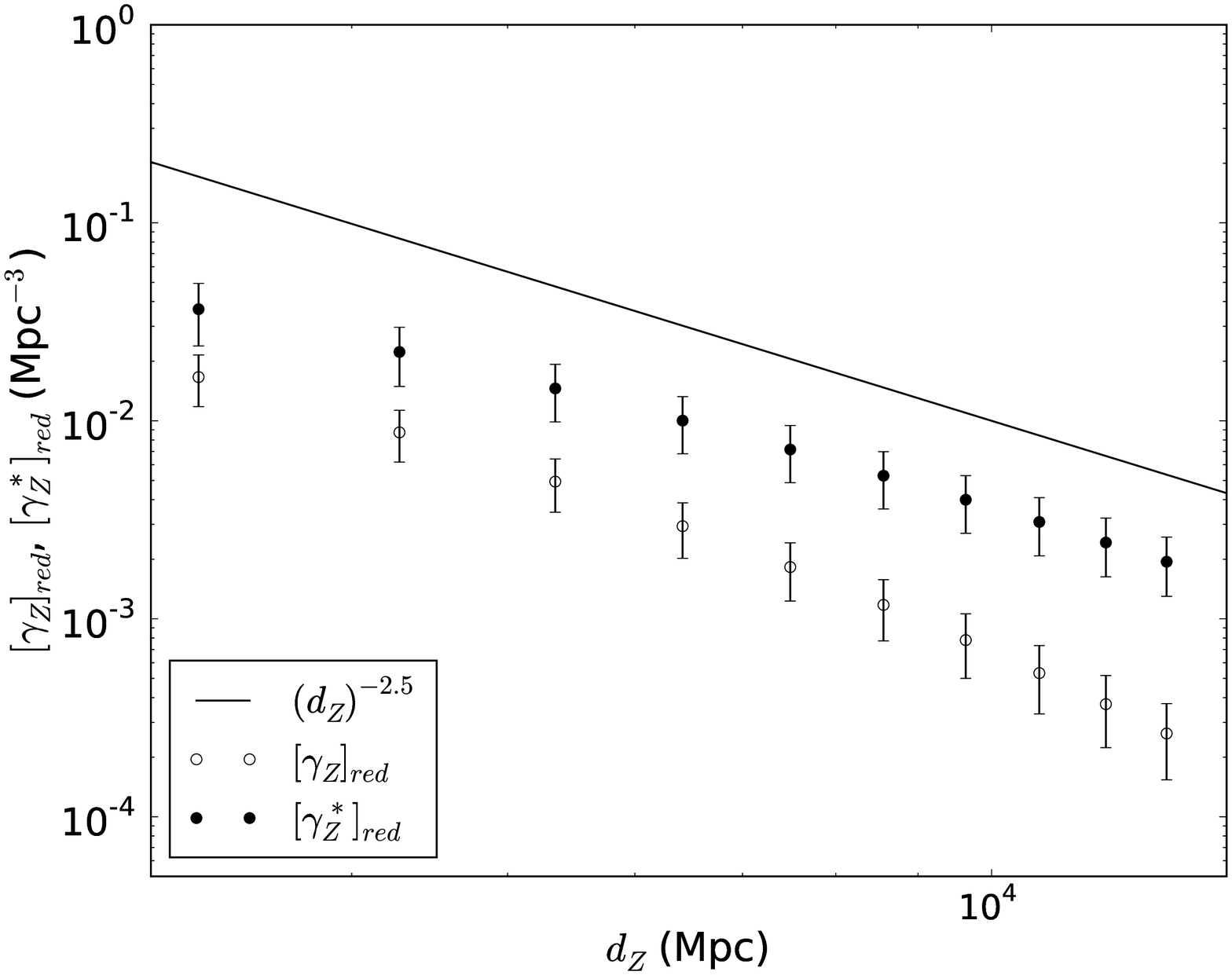}
\caption{{Observational} relativistic differential and integral
densities versus redshift distance in the combined red bands of the
FDF dataset of G06. The solid line is plotted as a reference to the
power law behavior stated in the legend. Symbols are as in the 
legend.}
\label{reddzplot}
\end{figure}

A similar pattern can be found in figures \ref{UVdzplot}, \ref{Optdzplot},
and \ref{reddzplot}, which plot the observational differential and
integral densities using the redshift distance against that same
distance in the combined UV, optical and red bands respectively.
We found a similar indication of a power-law pattern in both
densities at high redshifts, but the actual power exponent of the
distributions is different from the one using the luminosity
distance. This comes from the fact that these two relativistic
distances are affected differently by the expansion of the
FLRW spacetime. Comparing the results shown in figure
\ref{Optdzplot} with the ones in the UV band in figure \ref{UVdzplot}
we can see that the slopes of the data points should agree with
one another. Similarly as in figures \ref{UVdzplot} and \ref{Optdzplot},
the power law exponent shown in figure \ref{reddzplot} seems to be
the same for both $\obs{\gamma_{\ssty Z}}$ and
$\obs{\gamma^\ast_{\ssty Z}}$ in the given distance definition.

One should notice that the slope of the straight lines
used as references for the data points in figures \ref{UVdlplot} to
\ref{reddzplot} do not vary drastically over all bands of the FDF
datasets, both for luminosity and redshift distance definitions.
This possibly indicates that such an effect is predominantly
relativistic.

The differential density constructed with the area distance
$\da$ becomes discontinuous at $z \approx 1.5$ due to the fact that,
by definition, $\da$ has a maximum at that redshift and, therefore,
its derivative with respect to $z$ becomes zero, yielding
an undefined $\obs{\gamma_{\ssty A}}$ at high redshifts. In addition,
differently from the decrease of $\gamma^\ast_{\ssty L}$ and $\gamma^\ast_{\ssty Z}$
at higher redshifts, the integral density
$\gamma^\ast_{\ssty A}$ increases with $z$. Such pathologies have
already been previously detected (see figs.\ 5 and 6 of A07). They
seem to render the  number densities defined by $\da$ as unphysical. 

The differential and integral densities defined by $\dg$
lack the geometrical effect of expansion in the particular
spacetime under consideration in this paper and, therefore, are
understood to be unsuitable if one wants to fully characterize the 
relativistic power-law patterns arising from the combination of
both spacetime expansion and LF evolution.

Finally, it is apparent that the differential densities
$\obs{\gamma_{\ssty L}}$ and $\obs{\gamma_{\ssty z}}$ do not appear
to exhibit a power-law decay as the integral ones do, especially at
redshift values nearing the sample's depth. This is very similar to
what found in A07. This behavior is
possibly a result of noisier data at the redshift limit of the
samples. By definition, the differential densities measure the rate
of growth in number counts, rendering them more sensitive to local
fluctuations, whereas the integral densities indicate the change in
number counts for entire observational volumes, rendering them
less sensitive to the same fluctuations.

The indication of power-law behavior may be the result of
observational, not necessarily spatial, inhomogeneity as we look
down our past light cone, combined with the incompleteness of
galaxy counts at higher redshifts. It may also be a result of
other causes which we are investigating.

\section{Conclusion} \lb{conclusions}

In this paper we have detailed a generalization of the framework 
connecting the relativistic cosmology number count theory with
the astronomical data extracted from the galaxy luminosity
function (LF), as proposed by Ribeiro \& Stoeger (2003).
This framework was used by Albani \etal (2007) to extract
the observed differential number counts from the LF data and to
study the observational inhomogeneities in the relativistic
radial densities using different distance definitions. Here we
have carried out {an analysis} of number counts using the
LF dataset of Gabasch \etal (2004) in the redshift range
$0.5\le z \le5.0$, and of Gabasch \etal (2006) in the
redshift range $0.75 \le z \le 3.0$. In doing so we focused on
the observational inhomogeneities in the relativistic radial
statistics of the distribution of the galaxies by means of the
differential densities $\gamma_{\ssty i}$ ($i = {\sty A}$, ${\sty G}$,
${\sty L}$, ${\sty Z}$) and integral densities $\gamma^{\ast}_{\ssty i}$,
both using the various cosmological distance definitions $d_i$,
as well as the selection functions and the differential
number counts $\obs{dN/dz}$ obtained from the LF of G04 and G06.

We confirmed the dependence of such empirically based relativistic
densities on the distance definition, a geometrical consequence of
the expansion of the Universe, which also is the cause of the
observational  inhomogeneities present even in theoretical
{analysis} of the spatially homogeneous FLRW spacetime.
Furthermore, as expected, the effect of the redshift evolution of
the LF seems to increase the observational inhomogeneity of these
densities.
 
We found evidence of a power-law pattern in the behavior of
$\obs{\gamma_{\ssty L}}$ and $\obs{\gamma_{\ssty z}}$,
relative to their respective cosmological distance measures
$\dl$ and $\dg$, similar to that previously detected in A07.
However, this pattern is somewhat less pronounced than the
results of A07. This difference could be due to the
use of a different, and somewhat more complete, redshift samples
used in this paper, especially the red galaxies of G06.

Finally, we should note the dependence of these results and
their interpretation on the assumed FLRW cosmology. Therefore,
we look forward to finding out what an analysis of this data
with reference to an inhomogeneous cosmological model (e.g.,
a Lema\^{\i}tre-Tolman-Bondi model) would reveal. If the
universe is not almost-FLRW or if the data are not probing
almost-FLRW scales, then hidden in the results are the effects
of having chosen the wrong cosmological model. In addition, what
we interpret as LF evolution, which is the same as number
evolution, but with some indirect luminosity evolution effects,
undoubtedly also reflects a lot of unobserved, and therefore
uncounted, galaxies. Strictly speaking, if the differential and
integral densities are comoving densities, then $\gamma_{\ssty G}$
and $\gamma^\ast_{\ssty G}$ should increase with redshift, as
there should be more galaxies in a comoving volume in the past
that there are now due to mergers (more, but smaller, galaxies
in the past). So we are obviously not observing as many galaxies
as we go farther out. Of course, that would mean that the average
mass per galaxy decreases with redshift, an effect that should be
implicit in the empirical LF, but which requires further
theoretical considerations to be made explicit in our model.
Including these points in our general approach to the problem
of connecting relativistic cosmology theory and the empirically
determined LF is indeed necessary in order to make the present
study even more realistic. We intend to deal with these issues,
as well as others like the possible meaning of power law patterns
in the differential and integral densities, in forthcoming papers.

Acknowledgments: Thanks go to A.\ Gabasch for kindly providing the 
absolute magnitudes of their I-band selected dataset and thoroughly 
helping us with its observational details. We are also grateful
to the referee for providing very helpful criticisms, recommendations
and remarks which substantially improved the paper. Finally, we
thank M.\ Giavalisco for some clarifying discussions.

\begin{center}
\begin{table}
\caption{Comparison with previous works  \label{novelties}}
\begin{tabular}{llll}
\hline	
Subject	       & RS03   &A07	&This work \\ \hline
redshift range & $0.05 \rightarrow 1.0$	&$0.05 \rightarrow 1.0$	&$0.5 \rightarrow 5.0$     \\
cosmology      & EdS	&FLRW	&FLRW      \\
classification & none	&none	&UV, optical, red  \\
methodology    & J implicit & FLRW implicit &J, FLRW clarified	\\
\hline
\end{tabular}
\end{table}
\end{center}

\begin{center}
\begin{table}
\caption{Selection functions for the FDF blue UV bands \label{selectUV}}
\begin{tabular}{lrlrlrl}
\hline
\multicolumn{1}{c}{redshift}
&\multicolumn{2}{c}{$\psi^{\ssty 1500}$}
&\multicolumn{2}{c}{$\psi^{\ssty 2800}$}
&\multicolumn{2}{c}{$\psi^{\ssty u'}$}	\\
\multicolumn{1}{c}{}
&\multicolumn{2}{c}{($10^{-3}$ Mpc$^{-3}$)}
&\multicolumn{2}{c}{($10^{-3}$ Mpc$^{-3}$)}
&\multicolumn{2}{c}{($10^{-3}$ Mpc$^{-3}$)}	\\
\hline
0.50	&31.0 &$^{+6.9}_{-5.9}$	&37.5	&$^{+7.9}_{-7.7}$	&42.7	&$^{+9.1}_{-7.9}$	\vspace{1 mm} \\
0.75	&16.4	&$^{+4.1}_{-3.6}$	&21.2	&$^{+4.9}_{-4.8}$	&25.2	&$^{+5.6}_{-5.0}$	\vspace{1 mm} \\
1.00	&9.6	&$^{+2.7}_{-2.4}$	&13.2	&$^{+3.4}_{-3.3}$	&16.2	&$^{+3.9}_{-3.5}$	\vspace{1 mm} 	\\
1.25	&6.1	&$^{+1.9}_{-1.7}$	&8.8	&$^{+2.5}_{-2.5}$	&11.1	&$^{+2.9}_{-2.6}$	\vspace{1 mm} 	\\
1.50	&4.1	&$^{+1.4}_{-1.3}$	&6.1	&$^{+1.9}_{-1.9}$	&8.0	&$^{+2.2}_{-2.1}$	\vspace{1 mm} 	\\
1.75	&2.8	&$^{+1.1}_{-1.0}$	&4.5	&$^{+1.5}_{-1.5}$	&5.9	&$^{+1.8}_{-1.7}$	\vspace{1 mm} 	\\
2.00	&2.05	&$^{+0.87}_{-0.80}$	&3.3	&$^{+1.3}_{-1.3}$	&4.5	&$^{+1.5}_{-1.4}$	\vspace{1 mm} 	\\
2.25	&1.53	&$^{+0.71}_{-0.65}$	&2.6	&$^{+1.0}_{-1.1}$	&3.5	&$^{+1.2}_{-1.2}$	\vspace{1 mm} 	\\
2.50	&1.16	&$^{+0.58}_{-0.54}$	&2.02	&$^{+0.88}_{-0.89}$	&2.80	&$^{+1.04}_{-0.99}$	\vspace{1 mm} 	\\
2.75	&0.91	&$^{+0.49}_{-0.46}$	&1.62	&$^{+0.76}_{-0.77}$	&2.26	&$^{+0.90}_{-0.85}$	\vspace{1 mm} 	\\
3.00	&0.72	&$^{+0.41}_{-0.39}$	&1.32	&$^{+0.65}_{-0.66}$	&1.85	&$^{+0.78}_{-0.74}$	\vspace{1 mm} 	\\
3.25	&0.58	&$^{+0.35}_{-0.33}$	&1.09	&$^{+0.57}_{-0.58}$	&1.54	&$^{+0.69}_{-0.65}$	\vspace{1 mm} 	\\
3.50	&0.48	&$^{+0.30}_{-0.29}$	&0.91	&$^{+0.50}_{-0.51}$	&1.29	&$^{+0.61}_{-0.58}$	\vspace{1 mm} 	\\
3.75	&0.40	&$^{+0.27}_{-0.25}$	&0.76	&$^{+0.45}_{-0.46}$	&1.09	&$^{+0.54}_{-0.51}$	\vspace{1 mm} 	\\
4.00	&0.33	&$^{+0.23}_{-0.22}$	&0.65	&$^{+0.40}_{-0.41}$	&0.93	&$^{+0.49}_{-0.46}$	\vspace{1 mm} 	\\
4.25	&0.28	&$^{+0.21}_{-0.20}$	&0.56	&$^{+0.36}_{-0.37}$	&0.80	&$^{+0.44}_{-0.41}$	\vspace{1 mm} 	\\
4.50	&0.24	&$^{+0.18}_{-0.17}$	&0.48	&$^{+0.32}_{-0.33}$	&0.70	&$^{+0.40}_{-0.37}$	\vspace{1 mm} 	\\
4.75	&0.21	&$^{+0.16}_{-0.16}$	&0.42	&$^{+0.29}_{-0.30}$	&0.61	&$^{+0.36}_{-0.34}$	\vspace{1 mm} 	\\
5.00	&0.18	&$^{+0.15}_{-0.14}$	&0.37	&$^{+0.27}_{-0.27}$	&0.53	&$^{+0.33}_{-0.31}$	\vspace{1 mm} 	\\
\hline
\end{tabular}
\end{table}
\end{center}

\begin{center}
\begin{table}
\caption{Selection functions for FDF blue optical bands \label{selectopt}}
\begin{tabular}{lrlrl}
\hline
\multicolumn{1}{c}{redshift}
&\multicolumn{2}{c}{$\psi^{\ssty g'}$}
&\multicolumn{2}{c}{$\psi^{\ssty B}$}	\\
\multicolumn{1}{c}{}
&\multicolumn{2}{c}{($10^{-3}$ Mpc$^{-3}$)}
&\multicolumn{2}{c}{($10^{-3}$ Mpc$^{-3}$)}	\\
\hline
0.50	&55	&$^{+14}_{-12}$	&53	&$^{+13}_{-11}$	\\
0.75	&32.7	&$^{+8.5}_{-7.7}$	&31.4	&$^{+7.7}_{-7.2}$	\vspace{1 mm} 	\\
1.00	&21.7	&$^{+6.0}_{-5.5}$	&20.7	&$^{+5.4}_{-5.1}$	\vspace{1 mm} 	\\
1.25	&15.3	&$^{+4.6}_{-4.2}$	&14.6	&$^{+4.0}_{-3.9}$	\vspace{1 mm} 	\\
1.50	&11.3	&$^{+3.6}_{-3.3}$	&10.7	&$^{+3.2}_{-3.1}$	\vspace{1 mm} 	\\
1.75	&8.6	&$^{+2.9}_{-2.7}$	&8.2	&$^{+2.6}_{-2.5}$	\vspace{1 mm} 	\\
2.00	&6.8	&$^{+2.5}_{-2.3}$	&6.3	&$^{+2.1}_{-2.1}$	\vspace{1 mm} 	\\
2.25	&5.4	&$^{+2.1}_{-1.9}$	&5.0	&$^{+1.8}_{-1.8}$	\vspace{1 mm} 	\\
2.50	&4.4	&$^{+1.8}_{-1.7}$	&4.1	&$^{+1.5}_{-1.6}$	\vspace{1 mm} 	\\
2.75	&3.6	&$^{+1.6}_{-1.5}$	&3.3	&$^{+1.3}_{-1.4}$	\vspace{1 mm} 	\\
3.00	&3.0	&$^{+1.4}_{-1.3}$	&2.8	&$^{+1.1}_{-1.2}$	\vspace{1 mm} 	\\
3.25	&2.5	&$^{+1.2}_{-1.2}$	&2.3	&$^{+1.0}_{-1.1}$	\vspace{1 mm} 	\\
3.50	&2.1	&$^{+1.1}_{-1.0}$	&1.96	&$^{+0.89}_{-0.95}$	\vspace{1 mm} 	\\
3.75	&1.84	&$^{+0.98}_{-0.92}$	&1.67	&$^{+0.80}_{-0.85}$	\vspace{1 mm} 	\\
4.00	&1.59	&$^{+0.88}_{-0.83}$	&1.43	&$^{+0.71}_{-0.76}$	\vspace{1 mm} 	\\
4.25	&1.37	&$^{+0.80}_{-0.76}$	&1.23	&$^{+0.64}_{-0.69}$	\vspace{1 mm} 	\\
4.50	&1.20	&$^{+0.73}_{-0.69}$	&1.07	&$^{+0.58}_{-0.63}$	\vspace{1 mm} 	\\
4.75	&1.05	&$^{+0.66}_{-0.63}$	&0.93	&$^{+0.53}_{-0.57}$	\vspace{1 mm} 	\\
5.00	&0.92	&$^{+0.61}_{-0.58}$	&0.81	&$^{+0.48}_{-0.52}$	\vspace{1 mm} 	\\
\hline
\end{tabular}
\end{table}
\end{center}

\begin{center}
\begin{table}
\caption{Selection functions for the FDF red bands \label{selectred}}
\begin{tabular}{lrlrlrl}
\hline
\multicolumn{1}{c}{redshift}
&\multicolumn{2}{c}{$\psi^{\ssty r'}$}
&\multicolumn{2}{c}{$\psi^{\ssty i'}$}
&\multicolumn{2}{c}{$\psi^{\ssty z'}$}	\\
\multicolumn{1}{c}{}
&\multicolumn{2}{c}{($10^{-3}$ Mpc$^{-3}$)}
&\multicolumn{2}{c}{($10^{-3}$ Mpc$^{-3}$)}
&\multicolumn{2}{c}{($10^{-3}$ Mpc$^{-3}$)}	\\
\hline
0.75	&31.6	&$^{+8.6}_{-8.6}$	&35.9	&$^{+9.8}_{-9.9}$	&44	&$^{+14}_{-14}$	\vspace{1 mm} 	\\
1.00	&22.4	&$^{+6.2}_{-6.2}$	&25.0	&$^{+6.8}_{-6.8}$	&30.2	&$^{+9.8}_{-9.7}$	\vspace{1 mm} 	\\
1.25	&16.8	&$^{+4.8}_{-4.8}$	&18.5	&$^{+5.1}_{-5.2}$	&22.1	&$^{+7.3}_{-7.3}$	\vspace{1 mm} 	\\
1.50	&13.1	&$^{+3.9}_{-3.9}$	&14.2	&$^{+4.1}_{-4.1}$	&16.9	&$^{+5.8}_{-5.8}$	\vspace{1 mm} 	\\
1.75	&10.5	&$^{+3.3}_{-3.3}$	&11.3	&$^{+3.4}_{-3.4}$	&13.4	&$^{+4.8}_{-4.8}$	\vspace{1 mm} 	\\
2.00	&8.6	&$^{+2.9}_{-2.8}$	&9.1	&$^{+2.9}_{-2.9}$	&10.8	&$^{+4.0}_{-4.1}$	\vspace{1 mm} 	\\
2.25	&7.2	&$^{+2.5}_{-2.5}$	&7.5	&$^{+2.5}_{-2.5}$	&8.9	&$^{+3.5}_{-3.5}$	\vspace{1 mm} 	\\
2.50	&6.1	&$^{+2.3}_{-2.2}$	&6.3	&$^{+2.2}_{-2.2}$	&7.5	&$^{+3.0}_{-3.1}$	\vspace{1 mm} 	\\
2.75	&5.2	&$^{+2.1}_{-2.0}$	&5.3	&$^{+1.9}_{-2.0}$	&6.3	&$^{+2.7}_{-2.7}$	\vspace{1 mm} 	\\
3.00	&4.5	&$^{+1.9}_{-1.8}$	&4.5	&$^{+1.7}_{-1.7}$	&5.4	&$^{+2.4}_{-2.4}$	\vspace{1 mm} 	\\
\hline
\end{tabular}
\end{table}
\end{center}

\begin{center}
\begin{table}
\caption{Differential number counts \label{dNdztable}}
\begin{tabular}{lrlrlrl}
\hline
\multicolumn{1}{c}{redshift}
&\multicolumn{2}{c}{$[dN/dz]_{\ssty UV}$}
&\multicolumn{2}{c}{$[dN/dz]_{\ssty opt}$}
&\multicolumn{2}{c}{$[dN/dz]_{\ssty red}$}	\\
\multicolumn{1}{c}{}
&\multicolumn{2}{c}{($10^{9}$)}
&\multicolumn{2}{c}{($10^{9}$)}
&\multicolumn{2}{c}{($10^{9}$)}	\\
\hline
0.50	&5.4	&$^{+1.2}_{-1.1}$	&7.9	&$^{+1.9}_{-1.7}$	&\multicolumn{2}{c}{---}	\vspace{1 mm} 	\\
0.75	&5.2	&$^{+1.2}_{-1.1}$	&8.0	&$^{+2.0}_{-1.8}$	&9.2	&$^{+2.7}_{-2.7}$	\vspace{1 mm} 	\\
1.00	&4.3	&$^{+1.1}_{-1.0}$	&7.1	&$^{+1.9}_{-1.8}$	&8.6	&$^{+2.5}_{-2.5}$	\vspace{1 mm} 	\\
1.25	&3.44	&$^{+0.96}_{-0.91}$	&5.9	&$^{+1.7}_{-1.6}$	&7.6	&$^{+2.3}_{-2.3}$	\vspace{1 mm} 	\\
1.50	&2.67	&$^{+0.82}_{-0.78}$	&4.9	&$^{+1.5}_{-1.4}$	&6.5	&$^{+2.0}_{-2.0}$	\vspace{1 mm} 	\\
1.75	&2.07	&$^{+0.69}_{-0.66}$	&3.9	&$^{+1.3}_{-1.2}$	&5.5	&$^{+1.8}_{-1.8}$	\vspace{1 mm} 	\\
2.00	&1.61	&$^{+0.58}_{-0.56}$	&3.2	&$^{+1.1}_{-1.1}$	&4.6	&$^{+1.6}_{-1.6}$	\vspace{1 mm} 	\\
2.25	&1.26	&$^{+0.49}_{-0.47}$	&2.58	&$^{+0.96}_{-0.93}$	&3.9	&$^{+1.4}_{-1.4}$	\vspace{1 mm} 	\\
2.50	&0.99	&$^{+0.41}_{-0.40}$	&2.10	&$^{+0.82}_{-0.81}$	&3.3	&$^{+1.2}_{-1.2}$	\vspace{1 mm} 	\\
2.75	&0.79	&$^{+0.35}_{-0.34}$	&1.71	&$^{+0.71}_{-0.70}$	&2.8	&$^{+1.1}_{-1.1}$	\vspace{1 mm} 	\\
3.00	&0.63	&$^{+0.30}_{-0.29}$	&1.41	&$^{+0.61}_{-0.61}$	&2.34	&$^{+0.97}_{-0.97}$	\vspace{1 mm} 	\\
3.25	&0.51	&$^{+0.26}_{-0.25}$	&1.16	&$^{+0.53}_{-0.53}$	&\multicolumn{2}{c}{---}	\vspace{1 mm} 	\\
3.50	&0.42	&$^{+0.22}_{-0.22}$	&0.96	&$^{+0.47}_{-0.46}$	&\multicolumn{2}{c}{---}	\vspace{1 mm} 	\\
3.75	&0.34	&$^{+0.19}_{-0.19}$	&0.80	&$^{+0.41}_{-0.41}$	&\multicolumn{2}{c}{---}	\vspace{1 mm} 	\\
4.00	&0.29	&$^{+0.17}_{-0.16}$	&0.67	&$^{+0.36}_{-0.36}$	&\multicolumn{2}{c}{---}	\vspace{1 mm} 	\\
4.25	&0.24	&$^{+0.15}_{-0.14}$	&0.57	&$^{+0.31}_{-0.32}$	&\multicolumn{2}{c}{---}	\vspace{1 mm} 	\\
4.50	&0.20	&$^{+0.13}_{-0.12}$	&0.48	&$^{+0.28}_{-0.28}$	&\multicolumn{2}{c}{---}	\vspace{1 mm} 	\\
4.75	&0.17	&$^{+0.11}_{-0.11}$	&0.41	&$^{+0.25}_{-0.25}$	&\multicolumn{2}{c}{---}	\vspace{1 mm} 	\\
5.00	&0.145	&$^{+0.100}_{-0.097}$	&0.35	&$^{+0.22}_{-0.22}$	&\multicolumn{2}{c}{---}	\vspace{1 mm} \\
\hline
\end{tabular}
\end{table}
\end{center}

\begin{center}
\begin{table}
\caption{Spacetime expansion vs. galaxy evolution effects on
relativistic homogeneity \label{ratios}}
\begin{tabular}{llll}
\hline
Ratio	&$z=1$	&$z=3$	&$z=5$	\\ \hline
1-($\gaml$/$\gamg$)&0.926&0.9942&0.9987	\\
1-($[\gamg]_{\ssty UV}$/$\gamg$)&0.969&0.9969&0.9991		\\
1-($[\gamg]_{\ssty opt}$/$\gamg$)&0.949	&0.9930	&0.9880		\\
1-($[\gamg]_{\ssty red}$/$\gamg$)&0.938	&0.9880	&\multicolumn{1}{c}{---}\\
\hline
\end{tabular}
\end{table}
\end{center}

\appendix
\section{Numerical algorithm} \lb{app}
{This appendix describes step by step how to apply the theoretical
and numerical methodologies described in sections \S \ref{theory},
\S \ref{nprob} and \S \ref{obs} in order to obtain the results
used in this paper. Steps 1 to 6 involve
the calculation of the theoretical quantities, as discussed in
\S \ref{theory} and \S \ref{nprob}. Steps 7 to 10 show how to obtain
the corresponding observational quantities for a given LF, as
discussed in \S \ref{obs}.

{\it Step 1 -- Coupled differential equations}. Obtain the scale factor
$S(r)$ and the cumulative number count $N(r)$ as a function of the
radial coordinate, or comoving distance $r$, by solving the coupled
equations (\ref{NvsR}) and (\ref{SvsR}). We used the fourth order
Runge-Kutta method to obtain values of $S(r)$ and $N(r)$ at fixed
$r$ increments, assuming as initial conditions $N(0)=0$ and $S(0)=1$;

{\it Step 2 -- Auxiliary quantities}. Obtain the redshift $z(r)$
for each increment of $r$ using equation (\ref{red}) and the
corresponding value of the scale factor $S(r)$. Do the same for the
area distance $\da(r)$ using equation (\ref{da}), the differential
number count $dN/dz(r)$ using equation (\ref{dNdz2}), and the
comoving volume number density $\nc(r)$ using equation (\ref{radn});

{\it Step 3 -- Cosmological distances}. Having the values of $z(r)$
and $\da(r)$ calculated for each $r$ increment, obtain the values
of the galaxy area distance $\dg(r)$, and the luminosity distance
$\dl(r)$ by direct application of equation (\ref{rec}), Etherington's
reciprocity law itself. The values of the redshift distance $\dz(r)$
can be easily obtained for each $r$ increment using equation (\ref{dz});

{\it Step 4 -- Derivatives of the distances with respect to redshift}.
For each $r$ increment, calculate the values of the derivatives with
respect to redshift of the distances obtained in the previous step
by means of equations (\ref{diffda2}), (\ref{diffdl2}) and
(\ref{diffdg2});

{\it Step 5 -- Theoretical differential densities}. Obtain the values
of the theoretical differential densities $\gen{\gamma}(r)$ for each
of the distance definitions $d_i$ ($i = {\sty A}, {\sty G}, {\sty L},
{\sty Z}$) calculated previously for each $r$ increment, by means of
equation (\ref{gamma});

{\it Step 6 -- Theoretical integral densities}. Calculate for each
distance $\gen{d}(r)$ entry its corresponding spherical volume
$\gen{V}(r)$, by means of equations (\ref{volume}), then use those
values, together with the ones for the cumulative number count
$N(r)$, to obtain the values of the integrated differential densities
$\gen{\gamma^{\ast}}(r)$, by means of equation (\ref{gest3});

{\it Step 7 -- Selection functions}. Select a bin containing a group
of redshift values in a certain interval $[z_{a \rightarrow b}]$, 
from a point $z_a$ to a point $z_b$, taken from the ones obtained
in step 2, and obtain the corresponding values of the selection
functions $\psi^{\ssty W}(z_{a \rightarrow b})$ for each filter $W$
using equation (\ref{sf}), with the appropriate values for the LF's
Schechter parameters $\phi^{\ast}(z_{a \rightarrow b})$, $M^{\ast}(z_{a
\rightarrow b})$, and $\alpha(z_{a \rightarrow b})$, and different
absolute magnitude limits $M^{\ssty W}_{lim}(z_{a \rightarrow b})$,
given by equation (\ref{mlim});

{\it Step 8 -- Completeness functions}. Calculate the values of the
completeness function $J^{\ssty W}(z_{a \rightarrow b})$ by means of
equation (\ref{Jdef2}) using the selection functions
$\psi^{\ssty W}(z_{a \rightarrow b})$ calculated in step 7 and the
corresponding values of the comoving volume number density
$\nc[z_{a \rightarrow b}(r)]$ obtained in step 2;

{\it Step 9 -- Differential and cumulative number counts}. Obtain
the observational differential number count
$\obs{dN/dz}^{\ssty W}(z_{a \rightarrow b})$ in each filter $W$ by
means of equation (\ref{Jdiffza}) and the observational cumulative
number count $\obs{N}^{\ssty W}(z_{a \rightarrow b})$, by means
of equation (\ref{obsN});

{\it Step 10 -- Observational differential and integral densities}.
Obtain the observational differential densities
$\obs{\gen{\gamma}}^{\ssty W}(z_{a \rightarrow b})$, and the
integrated differential densities
$\obs{\gen{\gamma^{\ast}}}^{\ssty W}(z_{a \rightarrow b})$, by
using the values of $\obs{dN/dz}^{\ssty W}(z_{a \rightarrow b})$
and $\obs{N}^{\ssty W}(z_{a \rightarrow b})$ in place of their
theoretical counterparts $dN/dz$ and $N$ in equations
(\ref{gamma}) and (\ref{gest3}), respectively.
}

\begin{thebibliography}{}
\bibitem{albani}{Albani, V.\ V.\ L., Iribarrem, A.\ S., Ribeiro, M.\
        B., and Stoeger, W.\ R.\ 2007, ApJ, 657, 760; astro-ph/0611032,
        (\textbf{A07})}
\bibitem{bell}{Bell, E.\ F., Mcintosh, D.\ H., Katz, N., and
        Weinberg, M.\ D.\ 2003, \apjs, 149, 289}
\bibitem{blanton}{Blanton, M.\ R., Hogg, D.\ W., Bahcall, N.\ A.,
        Brinkmann, J.\ \etal 2003, \apj, 592, 819}
\bibitem{bouwens}{Bouwens, R.\ J., Illingworth, G.\ D., Franx, M.,
        and Ford, H.\ 2007, \apj, 670, 928}
\bibitem{ellis71}{Ellis, G.\ F.\ R.\  1971, in General Relativity
        and Cosmology, ed.\ R.\ K.\ Sachs (Proc.\ Int.\ School
	Phys.\ ``Enrico Fermi''; New York: Academic Press); reprinted
	in Gen.\ Rel.\ Grav., 41, 581, 2009}
\bibitem{ellis07}{Ellis, G.\ F.\ R.\ 2007, Gen.\ Rel.\ Grav., 39,
        1047}
\bibitem{etherington}{Etherington, I.\ M.\ H.\ 1933, Phil.\ Mag.\
         ser.\ 7, 15, 761; reprinted in Gen.\ Rel.\ Grav. 39, 1055,
         2007}
\bibitem{fried}{Fried, J.\ W., von Kuhlmann, B., Meisenheimer, K.,
         Rix, H.-W.\ \etal 2001, \aap, 367, 788}
\bibitem{FORS-blue}{Gabasch, A., Bender, R., Seitz, S., Hopp, U.\
        \etal 2004, \aap, 421, 41 (\textbf{G04})}
\bibitem{FORS-red}{Gabasch, A., Hopp, U., Feulner, G., Bender, R.\
        \etal 2006, \aap, 448, 101 (\textbf{G06})}
\bibitem{cosmos}{Gabasch, A., Goranova, Y., Hopp, U., Noll, S.\
        \etal 2008, \mnras, 383, 1319}
\bibitem{heidt}{Heidt, J., Appenzeller, I., Bender, R., B\"{o}hm, A.\
         \etal 2001, The FORS Deep Field, Reviews
         in Modern Astronomy, Vol. 14, R.E.\ Schielicke (Ed.),
         Astronomische Gesellschaft}
\bibitem{CNOC2}{Lin, H., Yee, H.\ K.\ C., Carlberg, R.\ G.,
         Morris, S.\ L.\ \etal 1999, \apj, 518, 533}
\bibitem{ly}{Ly, C., Malkan, M.\ A., Kashikawa, N., Shimasaku,
         K.\ \etal 2007, \apj, 657, 738}
\bibitem{norman}{Norman, C., Ptak, A., Hornschemeier, A.,
         Hasinger, G.\ \etal 2004, \apj, 607, 721}
\bibitem{peacock}{Peacock, J.\ A.\ 1999, Cosmological Physics
        (Cambridge: Cambridge Univ. Press)}
\bibitem[2006]{pk06} Pleba\'{n}ski, J., \& Krasi\'{n}ski, A.\ 2006,
        An Introduction to General Relativity and Cosmology
	(Cambridge University Press)
\bibitem{poli}{Poli, F., Giallongo, E., Fontana, A., Menci, N.\
        \etal, \apjl, 593, L1}
\bibitem{pozzetti}{Pozzetti, L., Cimatti, A., Zamorani, G., Daddi, E.\
        \etal 2003, \aap, 402, 837}
\bibitem{juracy}{Rangel Lemos, L.\ J., and Ribeiro, M.\ B. 2008,
        \aap, 488, 55; arXiv:0805.3336}
\bibitem{rib92}{Ribeiro, M.\ B.\ 1992, \apj , 395, 29; arXiv:0807.0869}
\bibitem{rib95}{Ribeiro, M.\ B.\ 1995, \apj , 441, 477; astro-ph/9910145}
\bibitem{rib01}{Ribeiro, M.\ B.\ 2001, Gen.\ Rel.\ Grav., 33, 1699;
         astro-ph/0104181}
\bibitem{paper2}{Ribeiro, M.\ B.\ 2005, \aap, 429, 65; astro-ph/0408316}
\bibitem{paper1}{Ribeiro, M.\ B., and Stoeger, W.\ R.\ 2003,
        \apj, 592, 1; astro-ph/0304094, (\textbf{RS03})}
\bibitem{roos}{Roos, M., Introduction to Cosmology, Chichester:
         Wiley, 1994}
\bibitem{rudnick}{Rudnick, G., Rix, H.-W., Franx, M., Labb\'{e},
         I.\ \etal 2003, \apj, 599, 847}
\bibitem{schechter}{Schechter, P.\ 1976, \apj, 203, 297}
\bibitem{sg}{Sparke, L.S., and Gallagher, J.S. 2000, Galaxies in
        the Universe (Cambridge University Press)}
\bibitem{stoeger}{Stoeger, W.\ R., Ellis, G.\ F.\ R., and  Nel,
        S.\ D.\ 1992, Class. Q. Grav., 9, 509}
\bibitem{tzanavaris}{Tzanavaris, P., and Georgantopoulos, I. 2008,
        \aap, 480, 663}
\bibitem{wilmer}{Wilmer, C.\ N.\ A., Faber, S.\ M., Koo, D.\ C.,
        Weiner, B.\ J.\ \etal 2006, \apj, 647, 853}
\end{thebibliography}
\end{document}